\documentclass[longauth]{aa} 

\usepackage{graphicx}
\usepackage{txfonts}
\usepackage{color}
\usepackage{ulem}
\usepackage{natbib}

\begin{document}

\title{The Star Formation Rate Density and Dust Attenuation Evolution over 12~Gyr 
   with the VVDS Surveys
   \thanks{Based on observations collected at the European Organisation
    for Astronomical Research in the Southern Hemisphere, Chile under
    programs 072.A-0586 (GTO), 073.A-0647 (GTO) and 177.A-0837 (LP) at
    the Very Large Telescope, Paranal, and based on observations
    obtained with MegaPrime/MegaCam, a joint project of CFHT and
    CEA/DAPNIA, at the Canada-France-Hawaii Telescope (CFHT) which is
    operated by the National Research Council (NRC) of Canada, the
    Institut National des Science de l'Univers of the Centre National
    de la Recherche Scientifique (CNRS) of France, and the University
    of Hawaii. This work is based in part on data products produced at
    TERAPIX and the Canadian Astronomy Data Centre as part of the
    Canada-France-Hawaii Telescope Legacy Survey, a collaborative
    project of NRC and CNRS.}  }

\titlerunning{The CSFRD since 12~Gyr traced by the VVDS}

\author{
O.~Cucciati\inst{1,2}
\and L.~Tresse\inst{1}
\and O.~Ilbert\inst{1}
\and O.~Le~F\`evre\inst{1}
\and B.~Garilli\inst{1,3}
\and V.~Le~Brun\inst{1}
\and P.~Cassata\inst{4}
\and P.~Franzetti\inst{3}
\and D.~Maccagni\inst{1,3}
\and M.~Scodeggio\inst{3}
\and E.~Zucca\inst{5}
\and G.~Zamorani\inst{5}
\and S.~Bardelli\inst{5}
\and M.~Bolzonella\inst{5}
\and R.M.~Bielby\inst{6,7}
\and H.J.~McCracken\inst{7}
\and A.~Zanichelli\inst{8}
\and D.~Vergani\inst{5}
}

\authorrunning{O.~Cucciati, L.~Tresse, O.~Ilbert et al.}

\offprints{Olga Cucciati (cucciati@oats.inaf.it)}

\institute{Laboratoire d'Astrophysique de Marseille (UMR 6110), CNRS-Universit\'e de Provence, 38 rue Fr\'ed\'eric Joliot-Curie, F-13388 Marseille Cedex 13, France
\and INAF-Osservatorio Astronomico di Trieste, via Tiepolo 11, I-34143 Trieste, Italy
\and INAF-IASF, via Bassini 15, I-20133 Milano, Italy 
\and Department of Astronomy, University of Massachusetts, Amherst, MA 01003, USA
\and INAF-Osservatorio Astronomico di Bologna, via Ranzani 1, I-40127 Bologna, Italy 
\and Department of Physics, Durham University, South Road, Durham DH1 3LE
\and Institut d'Astrophysique de Paris (UMR 7095), CNRS-Universit\'e Pierre et Marie Curie, 98bis boulevard Arago, F-75014 Paris, France
\and IRA-INAF, via Gobetti 101, I-40129, Bologna, Italy 
}

\date{Received -; accepted -}
 
\abstract
  {}	
  {We investigate the global galaxy evolution over $\sim$12~Gyr
  ($0.05\leq z\leq 4.5$), from the far ultraviolet (FUV) luminosity
  function (LF), luminosity density (LD), and star formation rate
  density (SFRD), using the VIMOS-VLT Deep Survey (VVDS), a single
  deep galaxy redshift survey with a well controlled selection
  function.}
  {We combine the VVDS Deep ($17.5 \le I_{AB}\le 24.0)$ and Ultra-Deep
  ($23.00 \le i'_{AB} \le 24.75$) redshift surveys, totalizing
  $\sim11000$ galaxies, to estimate the rest-frame FUV LF and LD,
  using a wide wavelength range of deep photometry ($337 < \lambda <
  2310$~nm). We extract the dust attenuation of the FUV radiation,
  embedded in the well-constrained spectral energy distributions.  We
  then derive the dust-corrected SFRD.}
 {We find a constant and flat faint-end slope  $\alpha$ in the FUV LF
  at $z<1.7$. At $z>1.7$, we set $\alpha$ steepening with $(1+z)$. The
  absolute magnitude $M^{*}_{FUV}$ steadily brightens in the entire
  range $0<z<4.5$, and at $z>2$ it is on average brighter than in the
  literature, while $\phi^{*}$ is on average smaller. The evolution of
  our total LD shows a peak at $z\simeq2$, clearly present also when
  considering all sources of uncertainty. The SFRD history peaks as
  well at $z\simeq2$.  It first steadily rises by a factor of $\sim6$
  during 2~Gyr (from $z=4.5$ to $z=2$), and then decreases by a factor
  of $\sim12$ during 10~Gyr down to $z=0.05$. This peak is
  mainly produced by a similar peak within the population of galaxies with
  $-21.5 \leq M_{FUV} \leq -19.5$.  As times
  goes by, the total SFRD is dominated by fainter and fainter
  galaxies. The mean dust attenuation of the global galaxy population
  rises fast by 1~mag during 2~Gyr from $z\simeq4.5$ to $z\sim2$,
  reaches slowly its maximum at $z\simeq1$ ($A_{FUV}\simeq2.2$~mag),
  and then decreases by 1.1~mag during 7~Gyr down to $z\simeq0$.}
  {We have derived the cosmic SFRD history and the total dust amount
  in galaxies over a continuous period of $\sim12$~Gyr, using a single
  homogeneous spectroscopic redshift sample. The presence of a clear
  peak at $z\simeq2$ and a fast rise at $z>2$ of the SFRD is
  compelling for models of galaxy formation. This peak 
  is mainly produced by bright galaxies ($L \gtrsim L^{*}_{z=2}$),
  requiring that significant gas reservoirs still exist at this epoch
  and are probably replenished by cold accretion and wet mergers,
  while feedback or quenching processes are not yet strong enough to
  lower the SF. The dust attenuation maximum is reached $\sim$2~Gyr
  after the SFRD peak, implying a contribution from the
  intermediate-mass stars to the dust production at $z<2$.}


\maketitle


\section{Introduction}\label{introduction}

A robust determination of the star formation rate history (SFRH) is a
crucial element to understand galaxy evolution.  As time
goes by, the reservoirs of pristine gas are transformed into first
generation of stars. During galaxy evolution, feedback processes,
gas accretion, mergers, and environment are expected to affect star
formation, and to contribute as well to galaxy mass assembly. The SFH therefore
contains an imprint of the collective outcome of all processes which
shape galaxies along time.  The study of the SFH has been pioneered by
the CFRS \citep{lilly96},  on the evolution of the luminosity density
since $z\sim1$, and by \cite{madau1996}, on the SFR using the CFRS and
$z\sim3-4$ samples identified with HST.  Since then,
extensive new SFRD measurements have been compiled up to $z\simeq6$
\citep[see, for example,][]{hopkins2006}.

The Star Formation Rate Density (SFRD) is usually derived from a mean
luminosity density, defined as $ \mathrm{LD} =
\int^{\infty}_{\mathrm{0}} \phi(\mathrm{L})\ \mathrm{L}\ \mathrm{dL}$,
with $\phi(\mathrm L)$ being the Luminosity Function (LF) and $L$ a
luminosity related to the SFR.  At first sight, the SFRD is a simple
and powerful tool to investigate the cosmic star formation history.
The various data from different samples have persistently shown a
broad picture consistent within factors of $\sim$3, out to
high-redshifts ($z\simeq6$), showing a rise out to $z\simeq1$ and a
decline from $z\simeq3$, with an unclear evolution within $1\lesssim z
\lesssim 2.5$ because this redshift desert has remained difficult to
probe.  From the present day to $z\simeq1$, a steady rise of the SFRD
by one order of magnitude is firmly corroborated using various
calibrators of SFR - like far ultraviolet (FUV), far infrared (FIR),
H$\alpha$, radio - but the scatter amongst different measurements
remains large. Because of the number of uncertainties that remain
along the chain of transformations to be applied to galaxy counts and
luminosities to be converted into star formation rates, the exact
shape of the SFRD evolution still remains to be established. The
selection function of each galaxy sample, including the imaging
surveys depth and image quality or the redshift completeness, requires
a number of corrections to compute volume densities, and the complete
shape of the luminosity function remains speculative, particularly at
the faint end. Moreover, the transformation of luminosities to SFR
depends on assumptions on the conversion factors and on the amount of
dust attenuation, still open issues.

To be able to derive a coherent galaxy evolution model of galaxy
evolution, it is necessary to trace the SFRD evolution with the same
reference indicator throughout cosmic time, and within a deep galaxy
redshift survey with a simple, well defined and controlled selection
function.

The selection of spectroscopic targets may have a strong impact on the
type of galaxies studied and hence on the relevance of this sample to
the full galaxy population. A tight control of the selection function
is necessary to avoid the propagation of biases to some types of galaxies or
redshift ranges. Several methods are used to pre-select galaxies,
either using colour criteria tailored to different redshift ranges and
populations (e.g. LBG, BzK, radio, far-IR,...), or pure magnitude
selection.  A follow-up spectroscopic survey with
high success rate has then the advantage to provide a sample of
galaxies with known redshift and controlled uncertainties, with stars
and broad-line active galaxy nuclei clearly identified as 
pre-selection techniques are largely unable to fully discard them.

Using rest-frame ultraviolet galaxy luminosities has become a common
approach at high redshift, as it is applicable up to the highest
redshifts studied so far ($z\simeq 7$, e.g. \citealp{bouwens09_2z6}).
The non-ionising ultraviolet light (912-3000~\AA) is emitted by
relatively massive ($\ge3~M_{\odot}$), short-lived ($<3~10^{8}$~yr)
stars and it traces the SFR averaged over the last $\sim10^{8}$~yr
once it is corrected for dust attenuation \citep{kennicutt1998}. This
SFR estimator assumes a constant SFR over longer times than the very
massive stellar population ($>>15~M_{\odot}$) contributing to the
H$\alpha$ emission. As opposed to the H$\alpha$ instantaneous SFR
estimator, it does not disentangle whether the radiation is linked to
the creation of new stars ($<10^{7}$~yr) or to more evolved stars
($<10^{8}$~yr), and thus it is sensitive to ageing of star formation
regions \citep{calzetti2008}.  An ideal measurement of the
instantaneous SFR, not affected by uncertainties on dust, would be
based on the simultaneous use of H$\alpha$ and infrared emissivities,
since the absorbed ionising flux heats the dust which re-emits in the
infrared \citep{calzetti2008}. The difficulty is that it has not yet
been possible to assemble large, deep and statistically complete
samples for both estimators.

Here we aim to derive the SFRD evolution since $z\sim4.5$ using the
rest-frame UV luminosity density from the VIMOS VLT Deep Survey (VVDS)
samples `Deep' and `Ultra-Deep', totalizing $\sim11000$ galaxies with
spectroscopic redshifts.  We take advantage of being able to compute
rest-frame ultraviolet emissivities over a large and deep area of sky,
which enables to trace its evolution in a consistent way since
$z\simeq4.5$.

This paper is organised as follows. Section~2 gives a summary of our
VVDS data sets. Sections~3~\&~4 describe the ultraviolet luminosity
functions (LF) and densities (LD) from $z=0.05$ to $z=4.5$ as derived
from these data. Section~5 presents our results for the evolution of
the dust attenuation and the dust-corrected SFRD in the past
$\sim12$~Gyr.  Sect.~6 summarises and discusses our results. Technical
details are given in the Appendixes. Throughout this paper, we use the
AB flux normalisation, and we adopt the concordance cosmology
($\Omega_{m}$, $\Omega_{\Lambda}$, $h$)~=~(0.3, 0.7, 0.7), with which
the age of the Universe is 5.7,3.2,1.3~Gyr at $z=1,2,4.5$.


\section{Data}\label{data}

\subsection{The VVDS Deep \& Ultra-Deep spectroscopic surveys}
\label{spectro}

The VIMOS-VLT Deep Survey (VVDS) is a spectroscopic investigation of
distant sources, carried out with the high-multiplex, wide-field
(224~arcmin${^2}$) VIsible Multi-Object Spectrograph
\citep[VIMOS,][]{lefevre2003} mounted at the Nasmyth focus of MELIPAL,
the third of the four 8.2m ESO-VLT Unit Telescopes in Paranal, Chile.
The VVDS is composed of three $I$-selected surveys totalizing about
47000 spectra of galaxies, quasars, and stars; (1) a Wide survey
\citep[$17.5 \le I_{AB}\le 22.5$][]{garilli2008}, (2) a Deep survey
\citep[$17.5 \le I_{AB}\le 24.0$][]{lefevre2004,lefevre2005a}, and (3)
an Ultra-Deep survey ($23.00 \le i'_{AB}\le24.75$, Le~F\`evre et al.,
in prep.).  It spans a wide redshift range; from $z>0$ to $z\simeq5$
for targeted sources, and up to $z\simeq6.6$ for serendipitous
Ly$\alpha$ emitters \citep{cassata2011}. The VVDS surveys are purely
flux limited, and are free from any colour pre-selection or
galaxy-star separation when preparing the target catalogues. In the
Deep survey, the projection of galaxy size on the x-axis of the image
has been used to maximise the number of targets in each VIMOS pointing
(see \citealp{ilbert2005} for further details). We refer the reader to
\cite{mccracken2003} for a complete discussion on photometric
completeness as a function of $I_{AB}$ magnitude and surface
brightness. The conclusion of their analysis is that the VIMOS Deep
Imaging Survey ``is essentially free of selection effects until at
least $I_{AB}=25$''. Namely, considering the flux limit of the
  Ultra-Deep survey, $I_{AB}\le24.75$, our photometric catalogue is
  90\% and 70\% complete in surface brightness down to 24.5 and 25
  mag/arcsec$^2$, respectively.

In this study, we use both the Deep and Ultra-Deep surveys obtained in
the VVDS-0226-04 field. The Deep spectra were collected with
integration times of 16000~seconds using the LR-Red grism ($R=210$,
$5500<\lambda <9500$~\AA), over 2200~arcmin$^{2}$ of sky area. They
include the 9842 spectra described in \cite{lefevre2005a}, plus 2826
spectra acquired later with the same set-up (Le~F\`evre et al. in
prep.). With the Ultra-Deep survey we obtained repeated observations
of several $z\ge1.4$ targets from the Deep survey with assigned VVDS
quality flags of 0, 1, or 2 (i.e., with the lowest confidence level in
the spectroscopic identification) in order to assess their real
redshift distribution as detailed in Le~F\`evre et al. (in prep.). The
Ultra-Deep spectra were collected using both the LR-Red grism and the
LR-Blue grism ($3700<\lambda<6700$ \AA, $R=180$) with integration
times of 65000~seconds in each set-up. They consists of 1200 new
targets acquired over 576~arcmin$^{2}$ of sky area within the
2200~arcmin$^{2}$ sampled by the Deep survey. Our present work is
based on 10141 and 622 galaxy spectra at $0.05< z\le 4.50$ from the
Deep and Ultra-Deep surveys, respectively.  Those spectra have a
spectroscopic identification at a confidence level higher than
$\sim50$\%, 60\%, 81\%, 97\% and 99\% (corresponding to the VVDS
quality flags 1, 9, 2, 3 and 4).

We do not have a measured redshift for every source to a given
apparent magnitude limit in the observed field of view; the averaged
target sampling rates are $\sim24\%$ and $\sim4\%$ for the Deep
($17.5\leq I_{AB}\leq 24.0$) and Ultra-Deep ($23.00\leq i'_{AB}\leq
24.75$) surveys, respectively.  For each survey, we accurately
estimated the selection function accounting for the facts that (i) a
fraction of sources of the parent photometric catalogue was targeted
for spectroscopic observations; (ii) a fraction of the targeted
sources yields reliable redshifts. The computation of the selection
functions practically translates into weights applied to each galaxy
when computing the luminosity functions. In addition, in the Deep
survey, we used the results from our repeated observations, to obtain
a 100\% confidence level in the redshift measurement of lower quality
spectra.  The weighting schemes for the Deep and Ultra-Deep surveys
are independent, and we refer the reader to
Appendix~\ref{appendix_weights_deep} and
Appendix~\ref{appendix_weights_udeep} for a detailed description.

\subsection{The absolute magnitudes}
\label{abs_mag}

For the astrometric and photometric catalogues, we take the following
broadband imaging surveys over the VVDS-0226-04 field acquired at the
Canada-France-Hawaii Telescope (CFHT) using wide-field mosaic cameras:
the $BVI$ VVDS imaging survey \citep{mccracken2003,lefevre2004b} with
the CFHT-12K camera, the $u^{*}g'r'i'z'$ CFHT-Legacy Survey
(CFHTLS-D1/W1 field, T0005 release) with the MEGACAM camera and the
$JHK_{s}$ WIRDS survey \citep{bielby2011_WIRDS} with the WIRCAM camera.

To derive intrinsic luminosities, we use our Algorithm for Luminosity
Function \citep[ALF,][]{ilbert2005}, that integrates routines of the
code {\it Le
  Phare\footnote{http://www.cfht.hawaii.edu/$\sim$arnouts/LEPHARE/lephare.html}}. Absolute
magnitude measurements are optimised accounting for the full
information given by the above multi-band photometric data, in a way
which minimises the dependency on the templates chosen to fit the
observed colours \citep[see A.1 in][]{ilbert2005}.  We choose the
template library from \cite{BC03} modulated by the attenuation of the
intrinsic stellar continuum, $A(\lambda) = k(\lambda) E(B-V)$. We
consider a grid for the intrinsic colour excess of the continuum light
($E(B-V)$=0, 0.1, 0.2, 0.3, 0.4 and 0.5~mag), and an empirical
attenuation curve, $k(\lambda)$, of the form defined by
\cite{calzetti2000}.  We verified that the absolute magnitudes are
indeed mostly independent from the chosen template set, especially for
those derived in the rest-frame FUV, NUV, U and B photometric
bands. For this purpose, we computed them again using different
template sets, like PEGASE.2 templates \citep{fioc1997} or a mixed set
with \cite{BC03} and \cite{Polletta2007} templates, as used in
\cite{ilbert2009}.  This very weak dependency of the derived absolute
magnitudes on the adopted templates is mainly due to the fact that we
use $i)$ a wide ($337 < \lambda <2310$~nm) wavelength range with the
observed $u^{*}Bg'Vr'i'Iz'JHK_{s}$ photometric broadbands to obtain a
robust template fit, and $ii)$ the $u^{*}$-CFHTLS broadband filter
(our bluest photometric information, $337 < \lambda < 411$~nm) that
begins to sample the non-ionising ultraviolet continuum
($91.2-300$~nm) at $z>0.1$.  The $u^{*}$ photometric information is
deep enough to be complete down to our VVDS spectroscopic limiting
magnitudes, based on the primary selection from the $I$-band CFHT-12K
images ($725 < \lambda < 930$~nm).  This is the case at least up to
$z\simeq3.5$ when the Lyman-break feature is shifted towards our
redder photometric bands, i.e., $Bg'Vr'i'I$. The observed $u^{*}$-381
corresponds to the rest-frame NUV-250 at $z\simeq0.4$ and to the
rest-frame FUV-150 at $z\simeq1.5$.  It spans the NUV-250 band ($184
<\lambda <280$~nm) from $z=0.2$ to $z=1.2$ and the FUV-150 band ($135
<\lambda <175$~nm) from $z=0.9$ to $z=2$. At $z>2$, the rest-frame
FUV-150 and NUV-250 luminosities are covered by observations with
filters redder than $u^{*}$ (up to the J band at $z=4$ for the
NUV-250). Our rest-frame FUV absolute magnitudes are based on template
extrapolation only for $z<1$, but still we verified that our FUV-based
results agree with our NUV-based ones. As an example of these
consistency tests, in Appendix~\ref{LFallbands_app} we compare the
rest-frame NUV and FUV luminosity functions.


\section{The ultraviolet Deep \& Ultra-Deep luminosity functions}\label{lf}

In this work, we have primarily estimated the rest-frame FUV
luminosity functions and densities to derive the SFRD history.  The
recent SFR is traced by the intrinsic non-ionising ultraviolet stellar
continuum ($91.2-300$~nm) of galaxies (see Sect.~\ref{introduction}).
Within this UV range, the far UV radiation (FUV-150) is a better SFR
indicator than the near UV radiation (NUV-250), because the NUV is
contaminated by evolved stars, while the FUV is dominated by the radiation 
from new, massive, short-lived stars \citep[see, e.g.,][]{madau98}.

\begin{figure*}
\centering
\includegraphics[width=0.8\linewidth]{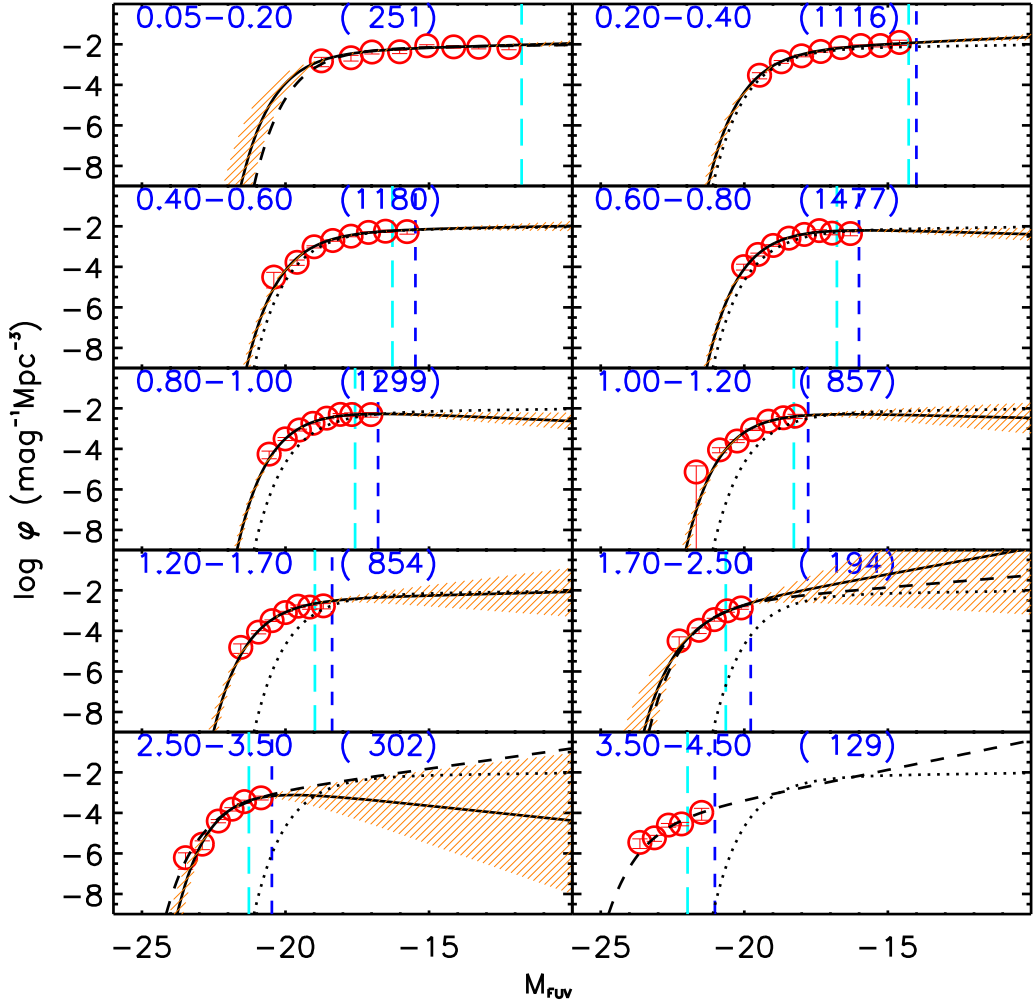}
\caption{Rest-frame FUV-band luminosity functions in ten redshift bins
  from $z=0.05$ to $z=4.50$ of the merged Deep+Ultra-Deep VVDS
  sample. Redshift ranges are indicated in each panel.  Red circles
  represent the $1/V_{max}$ data points with Poisson error bars, and the
  number of galaxies is given in parenthesis.  $1/V_{max}$ points are
  plotted up to the LF bias limit (see Sect.~\ref{lf_methods}) of the
  merged sample, represented with a vertical blue short-dashed
  line. This line corresponds also to the LF bias limit of the
  Ultra-Deep sample. The vertical cyan long-dashed line corresponds to
  the LF bias limit of the Deep survey. There is no Ultra-Deep data at
  $0.05<z\leq0.2$, so the only bias limit is the one of the Deep
  sample.  The black solid curve and its associated orange shaded area is
  the STY LF estimate assuming free Schechter parameters and its
  associated error.  At $3.5<z\leq4.5$, the STY LF fit does not
  converge using all free parameters.  In the first redshift bin, the
  dashed curve is the STY LF estimate when setting $M_{FUV}^{*}=-18.12$ (see text), 
  and for reference it is reported as a dotted curve in the other
  panels.  In the $1.7<z\leq2.5$, $2.5<z\leq3.5$, $3.5<z\leq4.5$
  panels, the dashed curve is the STY LF estimate when setting a
  faint-end slope that evolves with increasing redshift (i.e.,
  $\alpha=-1.3$, $-1.5$, $-1.73$, respectively).  The Schechter
  parameters of all the curves shown in this plot are listed in
  Table~\ref{FUVNUV_LF_table}, while Table~\ref{FUV_table_final}
  summarises those that we retained for our study.}
\label{FUV_LF_plot}
\end{figure*}

\begin{table*} 
\caption{LF Schechter parameters, LF absolute magnitude bias limits, LDs, SFRDs
and dust attenuation derived from our rest-frame FUV-150
LF estimates in 10 redshift bins from $z=0.05$ to $z=4.5$ assuming
($\Omega_{m}$, $\Omega_{\Lambda}$, $h$)~=~(0.3, 0.7, 0.7).}
\label{FUV_table_final} 
\centering 
\begin{tabular}{ccccccccc}
\hline\hline
{\tiny $\Delta z$}~\tablefootmark{a} & 
{\tiny $M^{*}_{FUV}$}~\tablefootmark{b} & 
{\tiny $\alpha$}~\tablefootmark{b} & 
{\tiny $\phi^{*}$}~\tablefootmark{b} & 
{\tiny $M_{Bias}^{D}$, $M_{Bias}^{UD}$}~\tablefootmark{c} & 
{\tiny lg(LD$_{uc}$)}~\tablefootmark{d} & 
{\tiny lg(SFRD$_{uc}$)}~\tablefootmark{e} & 
{\tiny $A_{FUV}$}~\tablefootmark{f} & 
{\tiny lg(SFRD$_{c}$)}~\tablefootmark{g}  \\ 
 &  
{\tiny AB~mag} &   
 & 
{\tiny $/$10$^{3}$Mpc$^{3}$}  & 
{\tiny AB~mag} & 
{\tiny W/Hz/Mpc$^{3}$} & 
{\tiny $M_{\odot}$/yr/Mpc$^{3}$} &  
{\tiny mag} & 
{\tiny $M_{\odot}$/yr/Mpc$^{3}$} \\         
\hline
0.05-0.2$\;\;$ &  $-18.12$		& $-1.05^{+0.04}_{-0.04}$    &    $7.00^{+0.44}_{-0.44}$	& $-11.8$, $...\; \; \; \; $   	& 18.76$^{+0.18}_{-0.18}$  	 &   $-2.09^{+0.18}_{-0.18}$  &  1.11 &  $-1.65^{+0.18}_{-0.18}$      \\
   0.2-0.4     &  $-18.3^{+0.1}_{-0.2}$	& $-1.17^{+0.05}_{-0.05}$    &    $6.91^{+1.02}_{-0.95}$	& $-14.3$, $-14.0$ 		& 18.87$^{+0.12}_{-0.12}$  	 &   $-1.98^{+0.12}_{-0.12}$  &  1.35 &  $-1.44^{+0.12}_{-0.12}$      \\
   0.4-0.6     &  $-18.4^{+0.1}_{-0.1}$	& $-1.07^{+0.07}_{-0.06}$    &    $6.60^{+0.91}_{-0.86}$	& $-16.3$, $-15.5$ 		& 18.85$^{+0.10}_{-0.10}$  	 &   $-2.00^{+0.10}_{-0.10}$  &  1.64 &  $-1.34^{+0.10}_{-0.10}$      \\
   0.6-0.8     &  $-18.3^{+0.1}_{-0.1}$	& $-0.90^{+0.08}_{-0.08}$    &    $9.53^{+0.99}_{-0.99}$	& $-16.8$, $-16.0$ 		& 18.93$^{+0.09}_{-0.09}$  	 &   $-1.92^{+0.09}_{-0.09}$  &  1.92 &  $-1.15^{+0.09}_{-0.09}$       \\
   0.8-1.0     &  $-18.7^{+0.1}_{-0.1}$	& $-0.85^{+0.10}_{-0.10}$    &    $9.01^{+0.94}_{-0.96}$	& $-17.6$, $-16.8$ 		& 19.04$^{+0.09}_{-0.08}$  	 &   $-1.79^{+0.09}_{-0.08}$  &  2.22 &  $-0.90^{+0.09}_{-0.08}$       \\
   1.0-1.2     &  $-19.0^{+0.2}_{-0.2}$	& $-0.91^{+0.16}_{-0.16}$    &    $7.43^{+1.08}_{-1.15}$	& $-18.3$, $-17.8$ 		& 19.12$^{+0.09}_{-0.09}$  	 &   $-1.74^{+0.09}_{-0.09}$  &  2.21 &  $-0.85^{+0.09}_{-0.09}$       \\
   1.2-1.7     &  $-19.6^{+0.2}_{-0.2}$	& $-1.09^{+0.23}_{-0.23}$    &    $4.10^{+0.77}_{-0.87}$	& $-19.0$, $-18.4$ 		& 19.13$^{+0.15}_{-0.08}$  	 &   $-1.72^{+0.15}_{-0.08}$  &  2.17 &  $-0.85^{+0.15}_{-0.08}$  \\
   1.7-2.5     &  $-20.4^{+0.1}_{-0.1}$	& $-1.30$ 		     &    $3.37^{+0.24}_{-0.24}$	& $-20.6$, $-19.8$ 		& 19.46$^{+0.49}_{-0.09}$  &   $-1.40^{+0.49}_{-0.09}$  &  1.94 &  $-0.62^{+0.49}_{-0.09}$    \\
   2.5-3.5     &  $-21.4^{+0.1}_{-0.1}$	& $-1.50$ 		     &    $0.86^{+0.05}_{-0.05}$	& $-21.3$, $-20.5$ 		& 19.40$^{+0.26}_{-0.15}$  &   $-1.45^{+0.26}_{-0.15}$  &  1.47 &  $-0.86^{+0.26}_{-0.15}$    \\
   3.5-4.5     &  $-22.2^{+0.2}_{-0.2}$	& $-1.73$		     &    $0.11^{+0.01}_{-0.01}$	& $-22.0$, $-21.2$ 		& 19.10$^{+0.22}_{-0.32}$  &   $-1.76^{+0.22}_{-0.32}$  &  0.97 &  $-1.37^{+0.22}_{-0.32}$    \\
\hline 
\end{tabular} 
\tablefoot{Values derived for the merged VVDS Deep+Ultra-Deep sample
  at $17.50\leq I_{AB}\leq 24.75$. There is no Ultra-Deep data at
  $z<0.2$.  \tablefoottext{a}Redshift bins.
\tablefoottext{b}Schechter parameters ($M^{*}$, $\alpha$ and
  $\phi^{*}$) of our VVDS rest-frame FUV-150 luminosity function that
  we use throughout this work.
\tablefoottext{c}Absolute magnitude
  limit in FUV down to which the survey is complete is terms of galaxy
  types (see Sect.~\ref{lf_methods}), for the Deep and the Ultra-Deep
  surveys, respectively.  
\tablefoottext{d}Rest-frame FUV luminosity
  density uncorrected for dust.  The quoted uncertainty includes errors
  on the LD induced by the STY LF fit, the Poisson noise, the cosmic
  variance, and the weighting scheme. When $\alpha$ is set at $z>1.7$,
  the STY errors include also the percentage uncertainty accounting for 
  the two extreme values of $\alpha$ ($-1.1$, $-1.73$) with respect to
  those quoted in the third column (see text). The relative
  contribution of each source of error is detailed in Table
  \ref{error_table_final} .
\tablefoottext{e}Star formation rate density uncorrected
  for dust, derived from the rest-frame FUV luminosity density. For 
  the errors see note (d).
\tablefoottext{f}Average dust attenuation in FUV.
\tablefoottext{g}Star formation rate density corrected for the dust
  attenuation $A_{FUV}$. For the errors see note (d).}  
\end{table*}

\subsection{The method}\label{lf_methods}

The galaxy luminosity function (LF) usually follows a
\citet{schechter1976} function characterised by a break luminosity,
$L^{*}$, a faint-end slope, $\alpha$, and a normalisation density
parameter, $\phi^{*}$. The LF is a fundamental measurement of the
statistical properties of the galaxy population; it is the
distribution of the galaxy comoving number density as a function of
their intrinsic luminosity at a given epoch. Despite its simple
definition, its estimation requires careful analyses of the survey
strategy, the selection criteria, and the completeness.  We derive it
using our code ALF that includes the non-parametric $1/V_{max}$, SWML,
and $C^{+}$ and the parametric STY luminosity function estimators
\citep[see Appendixes in][and references therein]{ilbert2005}.  Each
estimator presents advantages and drawbacks, and each one is affected
differently by different visibility limits for the various galaxy
types detected in deep flux-limited surveys.  In a given observed
band, galaxies are not equally visible to the same absolute magnitude
limit mainly due to the spectral type dependency of the $K$
corrections (i.e., different spectral energy distribution -SED- for
different spectral type galaxies).  Within a given redshift range, the
use of several non-parametric LF estimators allows us to empirically
determine the absolute magnitude range in which galaxies are equally
visible within our spectroscopic surveys.  The luminosity limit
down to which all galaxy populations are visible is called LF bias. As
discussed in \cite{ilbert2004}, $1/V_{max}$ and $C^{+}$ methods are
affected by this bias in a different way than the SWML and STY methods. If 
$1/V_{max}$ or $C^{+}$ at a given luminosity starts
giving different results from the SWML or STY, it means that a bias in
the global LF is present. The bias is significant if the two
estimators differ of more than the statistical uncertainties
(Poisson errors, in our case).  In each redshift bin explored, we
set the LF bias at the brightest absolute magnitude where $1/V_{max}$
or $C^{+}$ are different by more than 1-$\sigma$ from SWML or STY.  
Our LF parameters are estimated with data brighter than the LF bias
limit, to derive an unbiased LF faint-end slope of the global
population.  This is particularly important when combining surveys of
different depth. We use the \cite{schechter1976} functional form for
the STY estimate to calculate the Schechter parameters, because the
results are more robust than those with a simple fit with a Schechter
function of the non-parametric LF data points. The resulting faint-end
slope, $\alpha$, is independent on the luminosity binning, and since
the LF parameters are highly correlated to each other, we can account
for the allowed range for each Schechter parameter as derived by the
likelihood, in addition to the Poisson uncertainties typically quoted.

\subsection{The UV luminosity functions}
\label{lf_para}

We compute the rest-frame FUV LFs using a unique merged catalogue
which includes both the Deep and Ultra-Deep surveys, for a total
covered magnitude range of $17.5\leq I_{AB} \leq 24.75$. This way, we
exploit both the large magnitude range covered by the Deep survey and
the depth reached by the Ultra-Deep survey. This leads us to a robust
determination of the LF shape and normalisation.

When using flux-limited surveys with various apparent luminosity
depths, a coherent weighting scheme must take into account the
possible overlap of the flux ranges covered by the different surveys.
We detail the weights to be applied in this case in
Appendix~\ref{appendix_weights_merged}, and we verify their robustness
in Appendix~\ref{appendix_testing_weights}.  Fig.~\ref{FUV_LF_plot}
shows the rest-frame FUV-band LFs ($1/V_{max}$ and STY) obtained with
the merged catalogue with $17.50 \leq I_{AB} \leq 24.75$, in ten
independent redshift bins from $z=0.05$ to $z=4.5$. The non-parametric
$1/V_{max}$ data points are plotted up to the LF bias limit, that is
where those from the SWML and C$^{+}$ methods are in agreement (the latter
estimates are not shown throughout the paper for clarity in the
figures).  At $z<0.2$, due to the size of our survey fields, rare
bright nearby galaxies are not observed and the bright-end of the LF
cannot be constrained as shown by the large error (shaded area in
Fig.~\ref{FUV_LF_plot}) associated to the STY estimate.  For this
reason, in the first redshift bin we set $M^{*}$ to the local FUV
value, $M_{FUV}^{*}=-18.12$ \citep{wyder2005}.  Conversely, at
$z>1.7$, the faint-end of the LF starts to be loosely constrained.
Recent results in the literature (see Table~\ref{FUV_alpha_table})
seem to indicate a steep faint-end slope of the rest-frame FUV-band LF
at $z\gtrsim2$.  We therefore opt for a slope set to evolve with
increasing redshift, using the parameterisation, $\alpha(z)=A(1+z)+B$
\citep[see, e.g.,][]{ryan2007}.  To derive $A$ and $B$, we use our
highest redshift estimate of $\alpha$ ($\alpha=-1.1$ at $z\sim1.45$,
see Table~\ref{FUVNUV_LF_table}) and $\alpha=-1.73$ at $z=3.8$ taken
from \cite{bouwens2007}.  In this way, we obtain $\alpha=-1.3$,
$-1.5$, and $-1.73$ at $1.7<z\leq 2.5$, $2.5<z\leq 3.5$, and
$3.5<z\leq 4.5$, respectively.  In addition, in all the redshift bins
at $1.7<z\leq 4.5$ we compute $\phi^{*}$ and $M^{*}$ using also
extreme non-evolving $\alpha$ cases: $\alpha=-1.1$ as estimated at
$1.2<z\leq 1.7$, and $\alpha=-1.73$, as estimated at $z\sim 3.8$ by
\cite{bouwens2007}.  Table~\ref{FUVNUV_LF_table} summarises the
Schechter parameters for all above cases.

\begin{figure}
\centering
\includegraphics[width=0.9\linewidth]{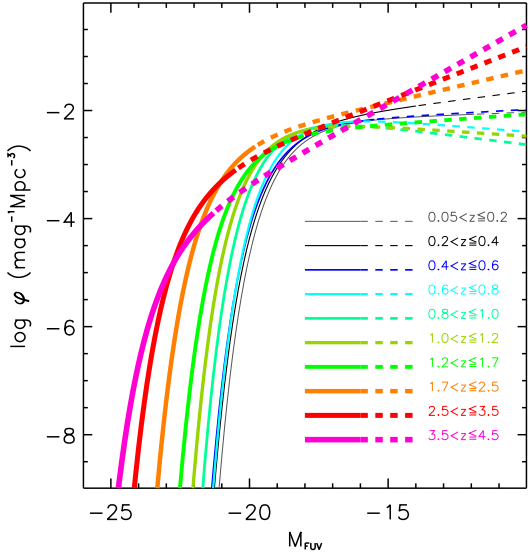}
\caption{The VVDS rest-frame FUV-band luminosity functions
  from $z\simeq0.1$ up to $z\simeq4$ (from the thinnest to the
  thickest curve), fitted with a Schechter functional form
  parametrised with the Schechter parameters ($\alpha$, $\phi^{*}$,
  $M^{*}$) reported in Table~\ref{FUV_table_final}. Lines are
  dashed for magnitudes fainter than the LF magnitude bias in each
  redshift bin (see Sec.~\ref{lf_methods}).}
\label{plot_LF_all}
\end{figure}

From now on, we consider our final {\it best} Schechter parameters ($\phi^{*}$,
$M^{*}$, $\alpha$) of the rest-frame FUV LFs as those estimated with the
STY method with $M^{*}$ set at the local value at $0.05<z\leq 0.2$, with all free
parameters at $0.2<z\leq 1.7$, and with $\alpha$ set to evolve at
$1.7<z \leq 4.5$.  They are reported in Table~\ref{FUV_table_final},
and the corresponding LFs are plotted all together in
Fig.~\ref{plot_LF_all}.  They are used to derive the luminosity
densities (Sect.~\ref{ld_fuv}) and the star formation rate densities
(Sect.~\ref{sfr_dust}). At $z>1.7$, the uncertainties of these
densities will include the possible span of $\alpha$ between the
values $-1.1$ and $-1.73$.

\begin{table} 
\caption{  Relative uncertainty to the total LD (sixth column of Table~\ref{FUV_table_final}) 
from different sources of error, expressed in percentage. Their sum in quadrature 
give the total error quoted in Table~\ref{FUV_table_final}. }
\label{error_table_final} 
\centering 
\begin{tabular}{c | l l l l l }
\hline\hline
 $\Delta z$~\tablefootmark{a} & STY~\tablefootmark{b} & $\alpha$ range~\tablefootmark{c} & Po~\tablefootmark{d}& CV~\tablefootmark{e}& $w$~\tablefootmark{f} \\ 
 & \multicolumn{5}{c}{errors in \%} \\         
\hline
0.05-0.2$\;\;$  & $^{+10.9}_{-9.8}$   & $-$ &  6.3  &  40	 &  $\sim$0	     \\
   0.2-0.4      & $^{+7.9}_{-6.5}$	     & $-$ &  3.0  &  26	 &  $\sim$0	     \\
   0.4-0.6      & $^{+5.0}_{-4.4}$	     & $-$ &  2.9  &  22	 &  $\sim$0	     \\
   0.6-0.8      & $^{+3.7}_{-3.3}$	     & $-$ &  2.6  &  20	 &  $\sim$0	     \\
   0.8-1.0      & $^{+3.9}_{-3.4}$	     & $-$ &  2.8  &  19	 &  $\sim$0	     \\
   1.0-1.2      & $^{+9.7}_{-6.4}$	     & $-$ &  3.4  &  19	 &  $\sim$0	     \\
   1.2-1.7      & $ ^{+32}_{-15}$    & $-$ &  3.4  &  12	 &  $\sim$0	     \\
   1.7-2.5      & $ ^{+5.5}_{-4.9}$    & $^{+112}_{-12}$    &  7.2  &  10	    &  $^{+11.3}_{-7.8}$    \\
   2.5-3.5      & $ ^{+3.5}_{-3.3}$    & $^{+56}_{-32}$    &  5.8  &  $\; \; 9$&  $^{+17.0}_{-9.6}$   \\
   3.5-4.5      & $ ^{+7.3}_{-6.4}$    & $^{+0}_{-55}$    &  8.8  &  10  &  $^{+48.4}_{-46.9}$  \\
\hline  	    
\end{tabular} 
\tablefoot{ For details on the computation of each error see Sec.~\ref{LD_results}. 
\tablefoottext{a}Redshift bins. 
\tablefoottext{b}Relative error induced by the STY LF fit. 
\tablefoottext{c}Relative uncertainty derived from the difference in the total LD between the `best' 
$\alpha$ value quoted in Table~\ref{FUV_table_final} and the lower ($\alpha=-1.1$) and upper 
($\alpha=-1.73$) limits considered. It is considered only when $\alpha$ 
is set. 
\tablefoottext{d}Relative error induced by Poisson noise.
\tablefoottext{e}Relative error induced by cosmic variance.
\tablefoottext{f}Relative error derived accounting for the error of the galaxy weighting 
scheme in the LF fit. }
\end{table}

\subsection{Main features of the derived FUV LF}
\label{LF_discussion}

Our results show a constant and flat ($\alpha\simeq -1$) FUV faint-end
slope at $z<2$, while we set it to steepen with $z$ at $z>2$, where
our data can not constrain it. We find that $M^*$ brightens
monotonically with increasing redshift for the entire redshift range
explored. Conversely, at $z\gtrsim 0.9$ $\phi^{*}$ starts decreasing,
with a remarkable drop especially at $z\gtrsim2$. These trends are
evident in Fig.~\ref{plot_LF_all} and in Table \ref{FUV_table_final},
and are also summarised in Fig.\ref{SFRD_picture}.

\subsubsection{A persistent flat faint-end slope at $z<2$}

Table~\ref{FUV_alpha_table} lists $\alpha$ values used in the
literature.  Contrarily to our work, at $z<2$ most previous studies
were not faint enough to determine the faint-end slope, which was then
often set to values as different as $-1.1<\alpha<-1.6$.  Nevertheless,
for instance, \cite{arnouts2005} and \cite{oesch2010_LF} estimated a
steep, but also constant, FUV faint-end slope $\alpha\simeq-1.6$ at
$0.4<z<1.2$ and $\alpha=-1.7$ at $0.75<z<2.5$, respectively. In
contrast, low-$z$ values ($z<0.4$) are found rather flat
$\alpha\simeq-1.2$ \citep[e.g.,][]{arnouts2005,wyder2005}.  To
establish the robustness of our results, Appendix~\ref{LFallbands_app}
details our three tests summarised below.

For the first test, we derived the rest-frame FUV $1/V_{max}$ estimates
using the VVDS photometric redshift catalogue, 1.25~mag deeper than
the VVDS spectroscopic catalogue (i.e., $I_{AB}=26$), over the same
VVDS area for a total of $\sim43000$ sources (see
Fig.~\ref{FUV_LF_plot_zphot}).  A similar flat faint-end slope is
found. Thus our derived flat slopes are not due to a large amount of
faint blue galaxies missing in the spectroscopic sample, or to an
inadequate weighting scheme. With this multi-wavelength ({\it
$u^{*}$g'r'i'z'}JHK$_{s}$) catalogue, we also tested the effect of
selecting galaxies upon their observed $u^{*}$-band up to 26~mag (i.e,
based on a rest-frame ultraviolet selection at $z<1$) rather than the
VVDS observed $I$-band selection (i.e., based on a rest-frame optical
selection at $z<1$). Even though the faint-end slope of such a sample
is slightly steeper, but not by more than $0.1$, it stays very
constant and close to flatness at $z\la 1$. Thus, our $I$-band
selection is unlikely missing a significant population of rest-frame
blue galaxies, that could have caused an hypothetic steeper slope.

In the second test, we confirm that the blue, or very blue galaxy
population of the VVDS $I$-selected spectroscopic survey, does not
give a steeper slope than the global galaxy population (see
Fig.~\ref{FUV_LF_plot_T4}).  In rest-frame bands redder than UV
(e.g. U, B), the shape of the LF is highly dependent upon the type of
galaxies. For instance, in \cite{zucca2006_VVDS_LF} at $0.4<z<0.9$, we
found $\alpha\simeq-0.9$(/$-1$) in $U$(/$B$)-band for the blue galaxy
population, $\alpha\simeq-1.7$(/$-1.6$) in $U$(/$B$)-band for the very
blue galaxy population, and $\alpha\simeq-1.2$ in both $U$ and $B$
bands for the global population.  The diversity of the shape occurs
because in bands redder than FUV, the radiation is contaminated by a
long-lived stellar population, that makes highly distinctive the
galaxy spectral energy distributions for different galaxy types. In
contrast, the FUV-band radiation is dominated by a single stellar
population of short-lived, massive stars, whatever the type of
galaxies.  This explains why we observe little change in our $\alpha$
FUV slope between the global galaxy population and the blue, or very
blue, galaxy populations.  Another possible additional effect is the
rising dust attenuation towards shorter wavelengths
\citep{driver2008}, that increases the dispersion of the rest-frame
$B-FUV$ colours. This dispersion, if dependent on the luminosity, may
weaken the $B$-band steep slope when using the same galaxy sample to
compute the LF in $FUV$-band.

In the third test, we derive the rest-frame NUV LFs at $0.05<z\leq
4.5$ (see Fig.~\ref{LF_NUV_app} and Table~\ref{FUVNUV_LF_table}).  The
faint-end slope is also flat at $z\la 1.5$ also in the NUV band,
showing that the flatness of our FUV slope is not due to possible
problems induced by our template extrapolations at $z<1$.

In summary, the robustness of our flat UV faint-end slope is
well-established at $z<2$, and it is more consistent with the
faint-end slope the local universe, i.e. $\alpha\simeq-1.2$
\citep{wyder2005}.  Despite its low value, we will see
(Sec. \ref{SFRD_dust_sec}) that our dust-corrected SFRD using our
rest-frame FUV LFs fully agrees with other measurements at $z<1$, a
cosmic period where the SFRD seems now well-constrained with different
SFR estimators.  This is because we correct the
FUV radiation for dust attenuation in an appropriate way as we describe later.

\subsubsection{Comparing FUV LFs at $z>2$}\label{LF_z2_discussion} 

At $z>2$, $M^{*}_{FUV}$ keeps on brightening, while $\phi^{*}$
drops. Thanks to the depth of our data, these trends are robust,
whichever the choice of $\alpha$ (see Table~\ref{FUVNUV_LF_table}).

We remark that the drop in $\phi^{*}$ that we see at $z>2$ is likely
unrelated to incompleteness of low surface brightness objects,
although this effect can not be completely ruled out. As mentioned in
Sec.~\ref{spectro}, the VVDS survey photometric catalogue, down
  to $I_{AB}\sim 24.75$, is 90(/70)\% complete in observed surface
  brightness down to 24.5(/25) mag/arcsec$^2$. As discussed in
\cite{mccracken2003} (Sec. 4.1), objects with particularly low surface
brightness, that could fall below our detection limit at high
redshift, are very few, while `normal' galaxies would be in any case
detected. Still, a possible concern can be raised if we consider
  that the above-mentioned observed surface brightness limits
  correspond to brighter and brighter absolute limits going to high
  redshift. We will see (Sec.~\ref{SFRD_lum}) that the major
  contribution to the total SFRD at $z>3$ is given by very luminous
  galaxies, so in principle our results at these epochs should be
  driven by galaxies with an absolute surface brightness bright enough
  to be within the detection limits. Nevertheless, we could miss a
  population of low absolute surface brightness galaxies, for which we
  would not be complete at high redshift.  We performed a very simple
  test to quantify, at a first order of magnitude, this possible
  effect. We considered the redshift bin $3.5<z<4.5$, where the I-band
  filter, used to select VVDS galaxies, corresponds to the rest-frame
  FUV light. This way, to pass from the surface brightness in I-band
  shown in \cite{mccracken2003} to an absolute FUV-band surface
  brightness we can neglect $K-$ and colour-corrections. We verified
  that all the galaxies brighter than the Deep\footnote{In this
    redshift range, only a very small fraction of the galaxies that we
    use to compute the LF are from the Ultra-Deep sample, as can be
    seen in Fig.~\ref{FUV_LF_plot_separated}.} LF bias ($M_{FUV}=-22$)
  have an absolute surface brightness brighter than 18 mag/arcsec$^2$,
  that coresponds, at $z\sim4$, to the observed limit of 25
  mag/arcsec$^2$ (considering only the cosmological dimming, with no
  $K-$ or colour-corrections), and only a small fraction is fainter
  than 17.5 mag/arcsec$^2$, corresponding at $z\sim4$ to the observed
  limit of 24.5 mag/arcsec$^2$. In summary, even if small effects can
  not be completely ruled out, we believe that our results are robust
  with respect to surface brightness completeness.

We also verified that the low $\phi^{*}$ value at $z>2$ is robust with
respect to the weighting scheme (see Appendix~\ref{appendix_weights})
that we apply to the spectroscopic sample. Thanks to this weighting
scheme, we are indeed able to properly recover the total galaxy
population in the photometric catalogue.  Fig.~\ref{FUV_LF_plot_zphot}
shows the FUV-band LFs computed using the entire photometric catalogue
(with photometric redshifts): it overlaps with the LFs computed using
the VVDS Deep+Ultra-Deep spectroscopic sample, reinforcing the
reliability of our weighting scheme.

Table~\ref{FUV_alpha_table} shows clearly that the determination of
$\alpha$ is a major source of uncertainty, and even when it is fitted
and not set, it can be as flat as $\sim -1.2$ or as steep as $\sim -1.7$.
The scatter on $\alpha$ is accompanied by analogous ones for
$M^{*}_{FUV}$ and $\phi^{*}$. We verified that our $M^{*}_{FUV}$ is
generally brighter than other works in literature, at all $z=2,3,4$,
while our $\phi^{*}$ is much smaller, especially at $z=3$ and 4 (see
e.g. \citealp{gabasch2004_LF} at $z=2,3$, \citealp{steidel99} at $z=3,4$, 
\citealp{arnouts2005} at $z=3$, \citealp{bouwens2007} at $z=4$,
\citealp{reddy_steidel2009} at $z=2,3$, \citealp{paltani2007} at
$z=3.5$, \citealp{gabasch2004_LF} at $z=3.5,4.5$,
\citealp{sawicky2006_LF} at $z=2,3,4$, \citealp{tresse2007} at
$z=3,3.6,4.3$). We remark, however, that in a few cases we are in
agreement with some previous work for what concerns $M^{*}_{FUV}$
and/or $\phi^{*}$: we agree on both $M^{*}_{FUV}$ and $\phi^{*}$ with
\cite{gabasch2004_LF} at $z=2.5$ and with \citealp{oesch2010_LF} at
$z\sim2$, and we agree on $M^{*}_{FUV}$ with \citealp{paltani2007} at
$z=3.5$ and \citealp{arnouts2005} at $z\sim3$.

This comparison cannot be exhaustive, because it is hard to directly
compare the LF shapes from different works. This happens for the
following reasons. First, the three Schechter parameters are linked
together; therefore, setting $\alpha$ will in some sense determine
also $M^{*}$ and $\phi^{*}$, and we already showed that $\alpha$ have
been set, or determined, to very different values. Second, computing
the Schechter parameters fitting the $1/V_{max}$ values or using the STY
method can lead to slightly different results. Third, different
FUV-band filters or central wavelength have been used in the
literature (peaked at $1350\leq \AA \leq 1700$, see
Table~\ref{FUV_alpha_table}).

As a final remark about comparing LFs in the literature, we detail in
Appendix~\ref{lf_counts_z3} the difficulty in comparing FUV luminosity
functions directly-observed (as ours) with those derived from FUV
number counts (as in \citealp{steidel99}, for instance). With this
analysis, we explain the following remarkable evidence: at $2.7\leq z
\leq 3.4$, on one side \cite{lefevre2005nat} and Le~F\`evre et al. (in
prep.) show that the VVDS presents surface number counts (per apparent
magnitude) at least two times larger than those found by
\cite{steidel99}, and this is particularly evident for bright galaxies
(I$_{AB}\lesssim 23.5$); on the other side, the VVDS rest-frame FUV LF
is only $\sim50$\% higher at bright magnitudes than the one found by
\cite{steidel99}. We refer the reader to Appendix~\ref{lf_counts_z3}
for details. The conclusion of this exercise is that number counts, in
colour-selected samples, are less representative of the complete
galaxy population than those in flux-limited samples, as the
respective $n(z)$ of these samples are very different. This exercise
also confirms that one can not easily transform number counts to LF in
the case of a skewed $n(z)$.


\section{The FUV comoving luminosity densities}\label{ld_fuv}

We derive the mean comoving luminosity density (LD) in each redshift
bin as $\mathrm{LD} = \int_{\mathrm{L_{faint}}}^{\mathrm{L_{bright}}}
\phi(\mathrm{L})\ \mathrm{L}\ \mathrm{dL}$, where $\phi(L)$ is the
luminosity function assuming a \cite{schechter1976} functional form as
done in Sect.~\ref{lf}. We set $L_{faint}=10^{15}$~W~Hz$^{-1}$ and
$L_{bright}=10^{25}$~W~Hz$^{-1}$ (corresponding to $M_{faint}\sim-3.4$
and $M_{bright}\sim-28.4$), to adopt the same limits as in the
compilation of \cite{hopkins2004} and to allow an easier comparison
with the literature. Given the typical shape of the LF (steeply
decreasing for bright galaxies), setting $L_{bright}$ as above or
$L_{bright}=\infty$ does not make any difference. Setting the faint
limit at $L_{faint}=10^{15}$~W~Hz$^{-1}$ gives a total LD on average
0.3\% different from the definition with $L_{faint}=0$.

\begin{figure*}
\centering
\includegraphics[width=0.8\linewidth]{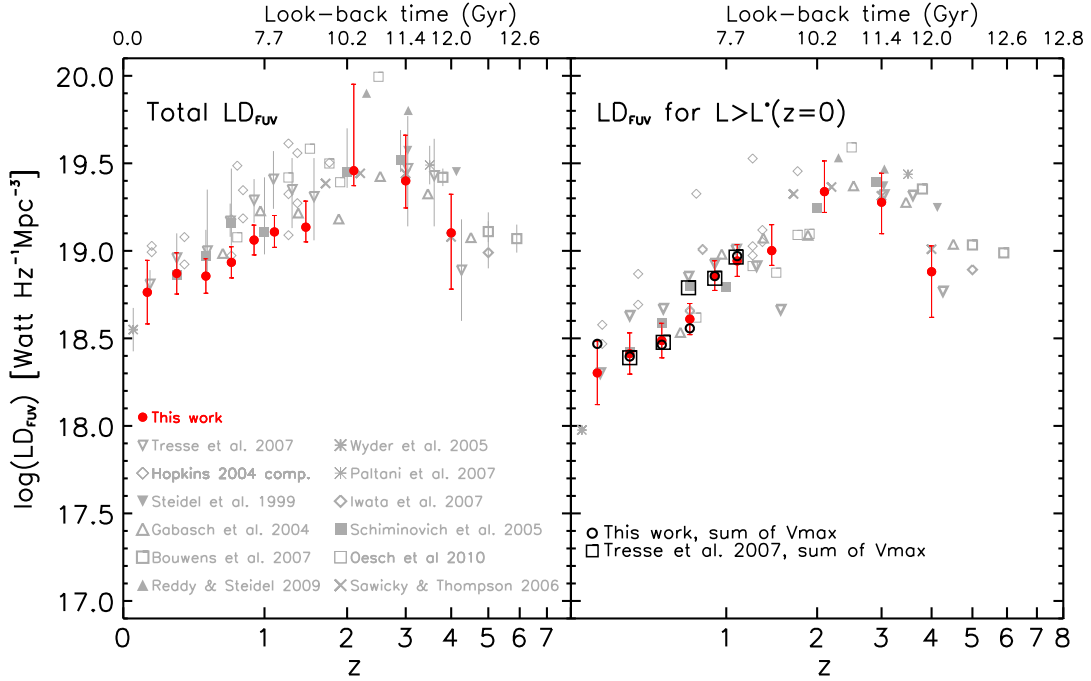} 
\caption{{\it Left panel}.  VVDS Deep+Ultra-Deep FUV-band luminosity
  densities derived from our {\it best} LF determination (FUV-150 red
  filled circles) as a function of redshift. Error bars are explained
  in Sect.~\ref{ld_fuv}. Gray data points are ultraviolet LF estimates
  found in the literature, as detailed in the labels (see also at the
  end of the caption).  {\it Right panel}. FUV-band luminosity
  densities restricted to galaxies brighter than $L^{*}$ at $z=0$ in
  \cite{wyder2005} ($M_{FUV}=-18.12$). Symbols are like in the
  left panel. Black open circles and squares represent the LD down to
  $M_{FUV}=-18.12$ summing the $1/V_{max}$ points in this work and in
  \cite{tresse2007}, respectively, instead of integrating the
  Schechter LF fit. We did this only up to $z\sim1$ because in VVDS
  data at $z\gtrsim1$ the bias limit for $1/V_{max}$ computation is
  brighter than $M_{FUV}=-18.12$. {\it List of references:}
  \cite{tresse2007} (VVDS Deep, FUV-150, empty upside-down triangles);
  compilation of FUV and NUV data extracted from \cite{hopkins2004}
  (thin open diamonds); \cite{steidel99} (FUV-1700, filled upside-down
  triangles); \cite{schiminovich2005} (FUV-150, filled squares those
  at $z<1.2$ GALEX-VVDS for $z<1.2$ data); \cite{bouwens2007}
  (FUV-1600, thick open squares); \cite{reddy_steidel2009} (FUV-1700,
  filled triangles); \cite{wyder2005} (GALEX-2dFGRS, FUV-150, thick
  asterisk); \cite{paltani2007} (FUV-1700, thin asterisk);
  \cite{iwata2007} (FUV-1700, thick open diamonds);
  \cite{gabasch2004_LF} (FUV-150, open triangles); \cite{oesch2010_LF}
  (FUV-150, thin open squares); \cite{sawicky2006_LF} (FUV-1700,
  crosses). }

\label{FUV_LD_plot}
\end{figure*}

\subsection{The VVDS FUV comoving luminosity density}
\label{LD_results}

The left panel of Fig.~\ref{FUV_LD_plot} shows the FUV-band LD that we
derived from our {\it best} LF determinations at $0<z<4.5$ and
tabulated with their total uncertainties in
Table~\ref{FUV_table_final}. Our LD uncertainties include errors from
the STY LF fit, errors due to cosmic variance, Poisson noise, and
errors associated to our weighting scheme. The relative
uncertainties induced on the total LD by each different source of
error are listed in Table~\ref{error_table_final}.

Errors derived from the STY estimator are underestimated at $z>1.7$
since $\alpha$ has been fixed in these redshift bins. In these cases,
we computed the LD also assuming the two extreme values of $\alpha$
(-1.1,-1.73), and the percentage difference with our {\it best} LDs at
$z>1.7$ was added in quadrature to the STY errors. Cosmic variance
(CV) errors are computed using the recipe in \cite{driver2010_cv}. We
computed CV errors also with the recipe in
\cite{trenti_stiavelli2008_cv}, and we found lower estimates at
$z<1.2$, going from $\sim20$\% at $z=0.1$ down to $\sim12$\% at
$z=1.1$. The same computation based on \cite{somerville2004_cv} gives
results similar to those by \cite{trenti_stiavelli2008_cv}. We decided
to use the more conservative values found with the recipe in
\cite{driver2010_cv}. Errors associated to the weighting scheme are
computed as follows. The weight for each galaxy has an error derived
in propagating the Poisson noise in the weighting formulas detailed in
Appendix~\ref{appendix_weights}. We computed again the LFs using the
weights plus and minus this error, and we derived the maximal and
minimal LDs. Their percentage difference with respect to the {\it
best} LDs gives errors arising from the weights. This error is
negligible below $z\sim1.7$.  In each redshift bin, the final LD
uncertainty is the addition in quadrature of all these sources of
errors. From Table~\ref{error_table_final}, it is evident that
CV errors dominate at $z<1.7$, while at $z>1.7$ the total uncertainty
is dominated by errors from the weighting scheme and by the percentage
difference when assuming the two extreme values of $\alpha$ with
respect to the chosen $\alpha$ value. It is worth noticing that our
largest uncertainty, the upper error bar at $z\sim2$, is given mainly
by the uncertainty on the value of $\alpha$: it includes the
difference between the LD with our {\it best} $\alpha$ ($=-1.3$) and
the extreme value $\alpha=-1.73$. We remind the reader that such
extreme $\alpha$ value, that we think is not optimal for our data at
this redshift, nevertheless has been found in the literature
\citep{reddy_steidel2009,oesch2010_LF} at very close redshift. Other
findings from other works span at this $z$ the range $-1.7\lesssim
\alpha \lesssim -1.1$. In Table~\ref{FUVNUV_LF_table} we give a
complete list of the LDs for our LFs described in
Sec.~\ref{lf_para}, i.e. including the {\it best} LFs, 
those with all free STY parameters at $z\sim0.1$ and $z>1.7$, and with
$\alpha$ set to $-1.1$ or $-1.73$ at $z>1.7$.

In the literature the treatment of cosmic variance as a source of
uncertainty differs from work to work. If we consider the data sets
overplotted in Figs.~\ref{FUV_LD_plot} and \ref{FUV_SFRDcorr_plot}, we
note the following. Some authors do not take into account the role of
cosmic variance (e.g., \citealp{schiminovich2005},
\citealp{wilson2002_SFH}, \citealp{iwata2007}); or consider it
negligible with respect to statistical uncertainties (e.g.,
\citealp{paltani2007}); or consider it negligible because of the large
volume explored (e.g, \citealp{vanderburg10_LF_CFHTLS}, 4 deg$^2$); or
mitigate its effects using several fields in different lines of
sights, although generally these fields are much smaller than the
VVDS-2h field (e.g., \citealp{steidel99},
\citealp{reddy_steidel2009}); or the cosmic variance has been included
as uncertainty in the LF normalisation (e.g.,
\citealp{bouwens2007}). We choose to compute it with the most
conservative method (see above) and to include it in our comprehensive
uncertainty on the LD.

The right panel of Fig.~\ref{FUV_LD_plot} shows the FUV-band LDs for
galaxies brighter than $L^{*}$ at $z\sim0$, i.e., the determination by
\cite{wyder2005}, corresponding to $M^{*}=-18.12$. In
Fig.~\ref{FUV_LD_plot}, we overplot data points of other studies which
published the three Schechter parameters of their rest-frame
ultraviolet LFs, that we integrated also to
$L_{faint}=10^{15}$~W~Hz$^{-1}$ (left panel) and to
$L_{faint}=L^{*}(z=0)$ (right panel).

From the left panel in Fig.~\ref{FUV_LD_plot}, it is evident that the
VVDS LD derived in this work shows a peak in its evolution with time.
After an increase by a factor of $\sim2$ from $z\sim4$ to $z\sim2$,
the LD decreases sharply by a factor of $\sim4.5$ down to $z\sim0$.
Fig.~\ref{plot_LF_all} shows that the LD is increasing from $z\sim0$
to $z\sim2$ because of the continuous brightening of $L^{*}$, with the
normalisation and the faint end slope not changing so much. The
brightening of $L^{*}$ also assures that the LD at $z\sim2$ would be 
robustly larger than at $z\sim1.5$, even if at $z\sim2$ we had used a flat
slope\footnote{We remind that the lower limits of the error bars at
$z>1.7$ include the possibility of $\alpha=-1.1$.}. In contrast, at $z>2$
the LD evolution is mainly driven by the evolution of the
normalisation: the LD decreases from $z\sim2$ to $z\sim4$ because of the
steep decrease of $\phi^{*}$. We note also that the decrease of the LD for
$z>2$ with our best estimate (red circles) is even smoothed away
because we are assuming $\alpha$ increasing with redshift (that should
cause the LD to increase). Only by using a much sharper steepening of
$\alpha$ at $z>2$ we could make the LD peak disappear, making the LD
staying roughly constant at $z>2$. For any other choice (the slope
slowly increasing with $z$, or fixed to a flat or steep value like in
the two extremes that we consider in the error bars), the LD is
clearly decreasing at $z>2$.

Our result is an improvement with respect to the similar study of the FUV-band LD in 
\cite{tresse2007}, that was based on our first VVDS-Deep spectroscopic sample
 \citep{lefevre2005a} and our UBVRI photometric sample
\citep{lefevre2004}.  On one side, the so-called redshift desert
($2<z<3$) could not be analysed, as it is now thanks to our Ultra-Deep
data.  On the other side, the spectroscopic and photometric data were
too shallow to determine $\alpha$ of the FUV-band LF, and thus a
constant value ($\alpha=-1.6$) was used over the explored redshift
range ($0<z<2$, $2.70<z<5$), as previously done in the literature (see
Table~\ref{FUV_alpha_table}). Thanks to our new spectroscopic and
photometric data, with the present work it is the first time that the
presence of a peak at $z\sim2$ in the FUV-band LD evolution has been
robustly assessed from a single survey.

\subsection{Discussion and comparison with literature}
\label{LD_discussion}

A large scatter between the different LD determinations is present in
both panels of Fig.~\ref{FUV_LD_plot}.  In integrating a Schechter LF
down to the faintest luminosities, the value of the faint-end slope
has a strong impact, especially if it is steep, leading to very
different LDs for different $\alpha$ \citep[see, e.g., Fig.~1
in][]{tresse2007}.  Table~\ref{FUV_alpha_table} shows that the
$\alpha$ determinations span a large range of values at every
redshifts, and it is evident that our determinations of $\alpha$ are
lower than the average, especially at $z<1.7$. This is the main
reason why our total LD appears on average lower than the other values
in the literature (but still in agreement with works that used $\alpha
\simeq -1$, like in \citealp{gabasch2004_LF}).  We have already
discussed in Sect.~\ref{lf} the difficulty to compare rest-frame FUV
LFs, and thus LDs, from various surveys, contrarily to rest-frame
optical LFs and LDs.  By integrating the LFs down to $M_{FUV}=-18.12$
(right panel of Fig.~\ref{FUV_LD_plot}), the differences in the values
of $\alpha$ should have little impact on the resulting LD, when this
integration limit is very close to $M^{*}$ (i.e. at $z<1$). The
observed scatter in the right panel of Fig.~\ref{FUV_LD_plot} means
that $L^{*}$ and $\phi^{*}$ are also discrepant among various data
sets.  Furthermore, since the three Schechter parameters ($L^*$,
$\phi^*$, $\alpha$) are strongly correlated, different values of
($L^{*}$, $\phi*$) arise even within the same data set when $\alpha$
is fixed to different values (see below). For instance, taking a
steeper $\alpha$ leads to a fainter $L^{*}$ and a lower $\phi{*}$ to
fit exactly the same data set.

In the right panel of Fig.~\ref{FUV_LD_plot} we also show our
$L>L^{*}_{z=0}$ LDs obtained by summing the $1/V_{max}$
values for galaxies with $L^{*}_{z=0} \leq L \leq L_{bright}$ (black
empty circles). The discrepancies between this estimate and the one
using the integration of the Schechter function mirror the level of
extrapolation made using a STY fit. We can see that, within the error
bars, our extrapolation is small, which ensures the goodness of our
fit at $M_{FUV}<-18.2$. The same exercise is repeated for the data in
\cite{tresse2007} (black empty squares).

We can test the effect of different LF parameterisations on total LD
estimates, using two works within the VVDS survey. \cite{tresse2007}
used the VVDS Deep data presented in \cite{lefevre2005a} ($UBVRI$) to
derive the rest-frame FUV-150 LDs. Their slope $\alpha$ has been set
to $-1.6$, based on GALEX-VVDS results from \cite{arnouts2005} at
$0.2<z<1$, since the depth of the U-band data available at that time
did not enable them to constrain the faint-end slope as we can do with
our new $u^{*}g'r'i'zJHK_{s}$ data.  Therefore, their LDs at $z\la
1.5$ are higher than those found in this work for both the total
galaxy sample and the $L>L^{*}_{z=0}$ galaxy sample. Nevertheless, if
one compares the minimal LDs (i.e., the sum of the $1/V_{max}$) between
this work and \cite{tresse2007}, they are in fairly good agreement.

Another example is given by the total LDs from
\cite{reddy_steidel2009} at $z\sim2$ and $\sim3$, that are a factor of
$\sim2.5$ larger than ours. The difference is mainly explained by the
much steeper slope in their LF (see Table~\ref{FUV_alpha_table}). In
fact, from the left panel of Fig.~\ref{FUV_LD_plot} we see that their
LD is in much better agreement with the upper limits of our error
bars, that correspond to the extreme case with $\alpha=-1.73$.

It is worth noticing that \cite{oesch2010_LF} find a constant $\alpha$
within $0.5<z<2.5$, but much steeper than ours
($\alpha\sim-1.7$). Moreover, they do find a peak in the LD evolution
as we do, but at $z\sim2.5$. We remark that they have to use points
from the literature to find this peak (the highest $z$ of their LD
determinations being $z=2.5$), while we are able to constrain the LD
peak using our own homogeneous data.

More generally, comparison with literature appears tricky, not only
because of very different assumptions on the LF shape (the need to set
$\alpha$ or $M^{*}$, for instance), but also because the LF shape can
be intrinsically different according to the various survey
characteristics (flux limit, band of selection...). Due to this
difficulty in comparing different works, it is clear the importance of
tracing the rest-frame FUV LDs over a wide range of redshifts with a
single and homogeneous survey. This is what we have been able to do
with our data set.


\section{The dust attenuation and the star formation rate densities up
to $z=4.5$}\label{sfr_dust}

A robust determination of the dust attenuation is a critical element
to transform the luminosity density into the effective star formation
rate density.  Thanks to our present VVDS Deep+Ultra-Deep
spectroscopic sample, combined with a wide range of deep
multi-wavelength data ($u^{*}g'r'i'zJHK_{s}$), we can now study in
details and with homogeneity the cosmic dust attenuation evolution
over $0<z<4.5$.

In the following subsections, we explain how we determined the dust
attenuation in our sample and how it compares with (and improves upon)
other measurements in the literature
(Sec.~\ref{dust_attenuation}). Then, we show how we derived the
dust-corrected SFRD for the global galaxy population
(Sec.~\ref{SFRD_dust_sec}), and how galaxies with different
luminosities contribute to
the total SFRD (Sec.~\ref{SFRD_lum}).

\subsection{The dust attenuation}
\label{dust_attenuation}

\begin{figure*} 
\centering
\includegraphics[width=0.8\linewidth]{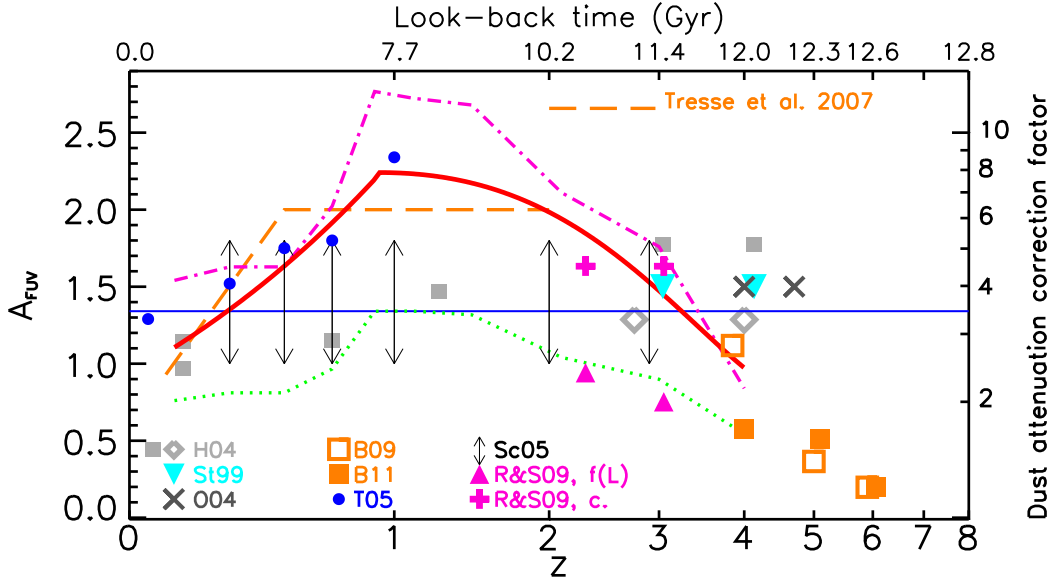} 
\caption{Dust attenuation $A_{FUV}$ in magnitudes as a function of
  redshift. The thick solid red curve represent the fit (see text for
  details) of the average dust attenuation, as a function of redshift,
  determined in this work, using the Calzetti's law. The $A_{FUV}$
  values derived from this fit in each redshift bin are listed in
  Table~\ref{FUV_table_final}. The magenta dot-dashed curve is the
  average $A_{FUV}$ determined in this work using the recipe in
  \cite{meurer1999_beta}, based on the $\beta$ slope. The green dotted
  curve is the same as the dot-dashed one, but here we use the recipe
  in \cite{cortese2006_dust}, calibrated with normal star forming
  galaxies.  The thin blue horizontal solid line is a constant
  $A_{FUV}$ computed with the Calzetti's law using one single typical
  value of $E(B-V)$ (=0.13). The right $y-axis$ shows the
  multiplicative factor to be applied to the observed luminosities,
  i.e., $10^{0.4 A_{FUV}}$. All the other symbols represent dust
  attenuations found in literature, as indicated in the labels: orange
  dashed line from \cite{tresse2007} (obtained comparing not-corrected
  SFRDs with dust-corrected SFRDs from other works, see text for
  details); gray open diamonds and filled squares from the compilation
  in \citet[H04]{hopkins2004} (only FUV and NUV determinations,
  diamonds and squares for a correction independent from or dependent
  on SFR, respectively); cyan filled upside-down triangles from
  \citet[St99]{steidel99}; thick gray crosses from
  \citet[O04]{ouchi2004}; orange open and filled squares from
  \citet[B09]{bouwens09_2z6} and \citet[B11]{bouwens11_beta4z7}
  respectively (slightly shifted in redshift for clarity); blue filled
  circles from \citet[T05]{takeuchi05_FUV_IR_dust}, based on the 02h
  field GALEX-VVDS LFs \citep{arnouts2005} and the CDFS field
  Spitzer 15-$\mu$m LFs \citep{leFloch05_IRLF} at
  $0.2<z<1$; arrows from \citet[Sc05]{schiminovich2005}, also based on GALEX-VVDS survey; filled
  magenta triangles and magenta crosses from
  \citet[R\&S09]{reddy_steidel2009} (triangles and crosses for a dust
  attenuation dependent on or independent from FUV luminosity,
  respectively). }
\label{AFUV_plot} 
\end{figure*}

Estimating the average dust attenuation properties in a galaxy
survey is a complex task. A number of authors have suggested various
methods to derive the dust attenuation using a multi-wavelength data
set. This leads to one important source of scatter in the literature
in the determination of the dust corrected SFRDs.  For instance, dust
attenuation can be estimated using dust reddening curves or UV slopes
in the FUV, or the Balmer decrement measured from H$\alpha$ and
H$\beta$ (with the drawback that H$\alpha$ is difficult to observe for
large samples at $z>0.3$). It can be also derived with
multi-wavelength SED fitting techniques, especially when the knowledge
of spectroscopic redshifts enables to break a dust-redshift
degeneracy.  Nevertheless, if there is a large and prominent
population of highly dust-obscured galaxies, it will be missed by deep
rest-frame FUV data sets. In principle, a powerful method is based on
adding the total SFR radiation, that is the one passing through dust
(FUV emission) and the one re-emitted by dust (FIR emission).
However, this method has 
potential drawbacks. For example, results will be exclusively linked
to those galaxies visible at both wavelengths, and mainly at $z<2$.
Furthermore, the situation in infrared survey observations is not optimal
due to unknown contents of cold and hot dust, or broad-line active
galaxy nuclei contamination, or large PSFs making difficult the
measurement of uncontaminated single sources. We refer the reader to
the discussion in  \cite{calzetti2008} for  details on computing dust
attenuation using both FIR and FUV data.

Since we want to derive the dust attenuation in a similar manner over
$0<z<4.5$, we estimate the dust attenuation using the results from the
SED fitting analysis. This method is not always fully reliable on a
object-by-object basis. Still, it provides a good estimate of the
average attenuation properties of the whole population under study and
of their evolution with time. Also it avoids the use of restricted
cross-matched optical-FIR catalogues.

We primarily use the recipe in \cite{calzetti2000} for
actively star-forming galaxies to derive our dust attenuations at
1500~\AA, and we also report and discuss those derived with other
recipes.  We follow the prescription in \cite{calzetti2000} (see
Eq.~4), that is, $A(\lambda)= E(B-V)_{star}\ k(\lambda)$, where
$E(B-V)_{star}$ is the intrinsic colour excess of the stellar
continuum of a galaxy, and $k(\lambda)$ is the starburst reddening
curve in \cite{calzetti2000}\footnote{$k(\lambda) =
2.659\times(-2.156+1.509/\lambda -0.198/\lambda^{2} +
0.011/\lambda^{3})+ 4.03$} at $[0.12-0.63] \mu$m.  As described in
Sect.~\ref{abs_mag}, our template SED fitting assigns to each galaxy a
value of $E(B-V)_{star}$ chosen from a grid of five possible values
($0.1, 0.2, 0.3, 0.4, 0.5$).  In each studied redshift bin, we compute
our resulting mean $E(B-V)_{star}$ and $A_{FUV}$.  We remark that our
$E(B-V)_{star}$ computations do not seem to depend on observed (i.e. not
dust-corrected) rest-frame FUV luminosities, which indicates that our
sample includes a large variety of actively star-forming galaxies at
each luminosity. For this reason, we are confident that our mean
$E(B-V)_{star}$ values, obtained averaging galaxies brighter than the
given magnitude bias at each redshift, do not depend on the range of
luminosity spanned at each redshift.

Fig.~\ref{AFUV_plot} shows our results. We find that $A_{FUV}$
increases steeply and fast from $z\sim4$ to $z\sim2$, i.e. of 1 mag
within only 2 Gyr. It keeps on mildly increasing from $z\sim2$ to
$z\sim1$ ($\sim0.2$ mag in 2.5 Gyr) and then it continuously decreases
by $\sim$1.1~mag within $\sim8$~Gyr from $z\sim1$ to $z\sim0$. This
results in a maximum dust attenuation value at $z\sim1$
($A_{FUV}\simeq 2.2$). We smooth $A_{FUV}(z)$ with a linear fit and a
$3^{rd}$ order polynomial function for the redshift ranges $z<1$ and
$z>1$, respectively, to preserve the different trends of $A_{FUV}(z)$
below and above this redshift. The red solid line in
Fig.~\ref{AFUV_plot} shows our smoothed $A_{FUV}$, and we give its
values in each redshift bin in Table~\ref{FUV_table_final}. We remark
that the chosen fits depart from the fitted data by a very small
amount, so we do not overplot the original $A_{FUV}$ for clarity.

We also compute our average $A_{FUV}$ with the recipe given by
\cite{meurer1999_beta}, based also on actively star-forming galaxies,
but using the $\beta$ slope of the UV continuum as a proxy to the UV
attenuation (dot-dashed magenta line in Fig.~\ref{AFUV_plot}). We use
the $\beta$ expression of \cite{kong2004_beta} which accounts for the
star formation history to reduce the large scatter of this
method. Using the $\beta$ slope leads to larger $A_{FUV}$, in
particular at $1\lesssim z \lesssim 2$. \cite{treyer2007} show that
this method works well at least up to $z=1$, but breaks down for
red-sequence and very blue compact galaxies and for the majority of
high-$z$ Lyman break galaxies, which form a low-attenuation sequence
of their own. Also, \cite{wijesinghe2011} emphasise that this method
overestimates SFRs unless a modified relation between $\beta$ and the
attenuation factor is used as indicated by other galactic properties
than the UV continuum, in particular for high-$z$ galaxy SFRs.
Therefore, a good knowledge of the galaxy population under study is
necessary for a reliable application of this method. For instance, we
compute our average values of $A_{FUV}$ with the $\beta$ expression
given by \cite{cortese2006_dust}, calibrated on normal star-forming
galaxies, and we plotted them in Fig.~\ref{AFUV_plot} as well (green
dotted line). While this relation follows the same trend as the
previous estimates, the resulting attenuations are lower by more than
1~mag.  On the basis of this analysis, we conclude that taking the
\cite{calzetti2000} reddening curve appears the best solution for the
VVDS Deep+Ultra-Deep sample since it is largely dominated by actively
star-forming galaxies (from $\sim40$\% at $z\sim0.1$ up to $\sim85$\%
at $z\sim4.0$, see Appendix~\ref{LFallbands_app}). Nevertheless,
whatever the method to compute the dust attenuation, the amount of
dust increases from $z=4$ to $z=1$, and then decreases down to
$z\sim0$.

In Fig.~\ref{AFUV_plot} we also overplot $A_{FUV}$ dust attenuation
values taken from the literature. We can see that data are diversely
spread between 1 and 1.8~mag. Actually, they are often derived in
different ways, depending sometimes on FUV-band luminosity
\citep{reddy_steidel2009}, or on SFR \citep{hopkins2004}, or being a
constant value \citep[][GALEX-VVDS]{schiminovich2005}. Conversely,
\cite{tresse2007} derived them by comparing the dust corrected
$12\mu$m \citep{perez_gonzalez2005} and H$\alpha$
\citep{tresse1998,tresse2002} SFRDs to their uncorrected FUV-derived
VVDS Deep SFRDs. They found that the attenuation at 1500~$\AA$ was
$\sim$2 mag from $z = 2$ to $z = 0.4$, and then it decreased from $z =
0.4$ to $z = 0$ down to $\sim$1 mag. These last $A_{FUV}$ values are
very close to our present work. We plot them in Fig.~\ref{AFUV_plot}
with an orange dashed line. We are in good agreement also with
\cite{takeuchi05_FUV_IR_dust} at $0.2<z<1$. To estimate the mean dust
attenuation, they compared the GALEX-VVDS LFs \citep{arnouts2005} to
the Chandra Deep Field South Spitzer 15-$\mu$ LFs
\citep{leFloch05_IRLF}. Our $A_{FUV}$ agrees also with the one derived
by \cite{steidel99} at $z\sim3$ and by \cite{bouwens09_2z6} at
$z\sim4$ (but see \citealp{bouwens11_beta4z7} for a much lower
$A_{FUV}$ value).

In summary, with the comprehensive VVDS Deep+Ultra-Deep sample we are
able to compute the evolution of the mean cosmic dust attenuation over
$\sim$12~Gyr.  It increases rapidly of 1 mag in 2 Gyr (from
$z\sim4$ to $z\sim2$), then it reaches a maximum at $z\sim1$
(increasing by $\sim0.2$ mag in 2.5 Gyr), and finally it decreases
continuously down to $z\sim0$ ($\sim$1.1~mag within $\sim8$~Gyr.) 
There is a maximum in dust attenuation at $z\sim1$ ($A_{FUV}\simeq
2.2$).

\subsection{The dust-corrected SFRD}\label{SFRD_dust_sec}

\begin{figure*} 
\centering
\includegraphics[width=0.7\linewidth]{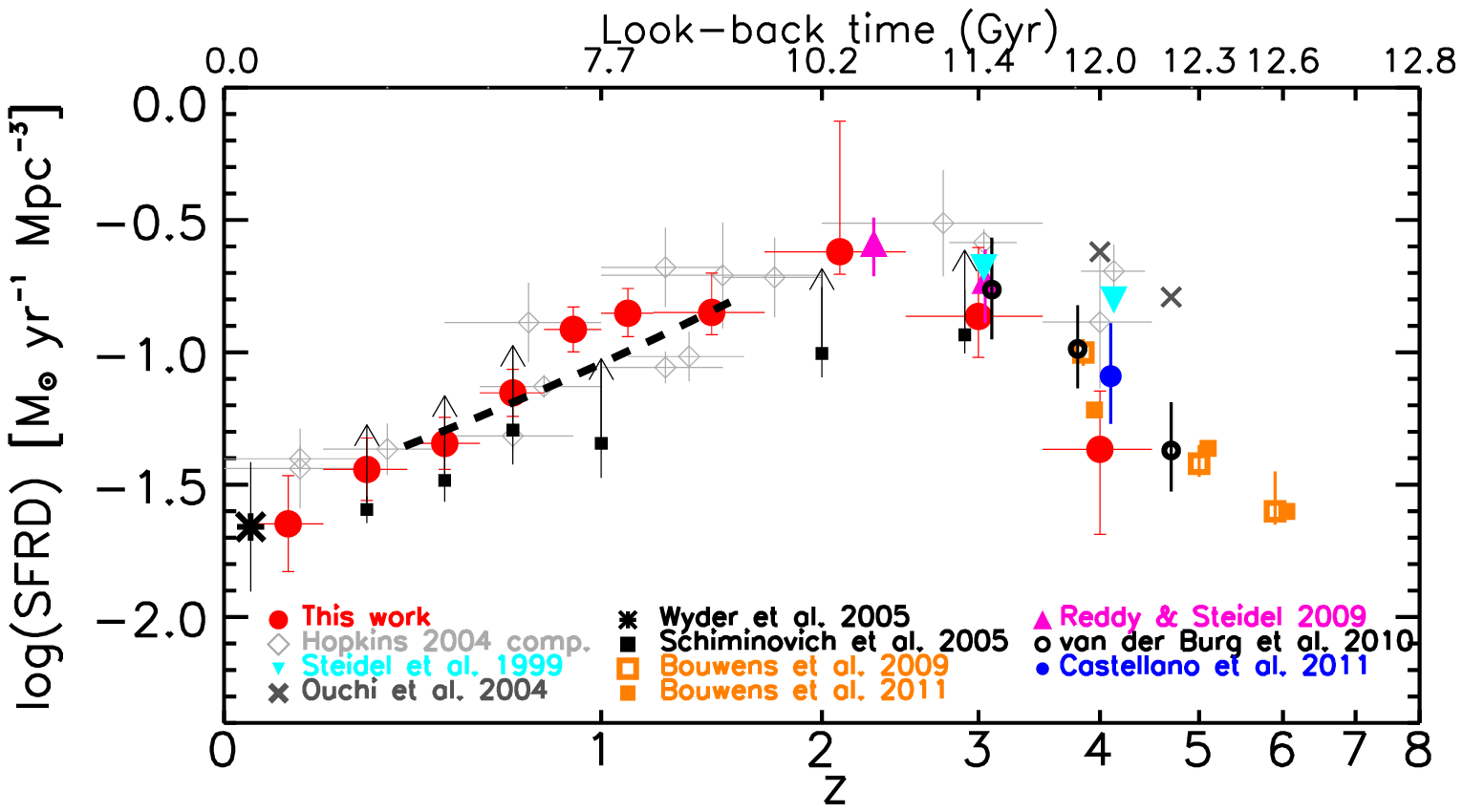}
\includegraphics[width=0.7\linewidth]{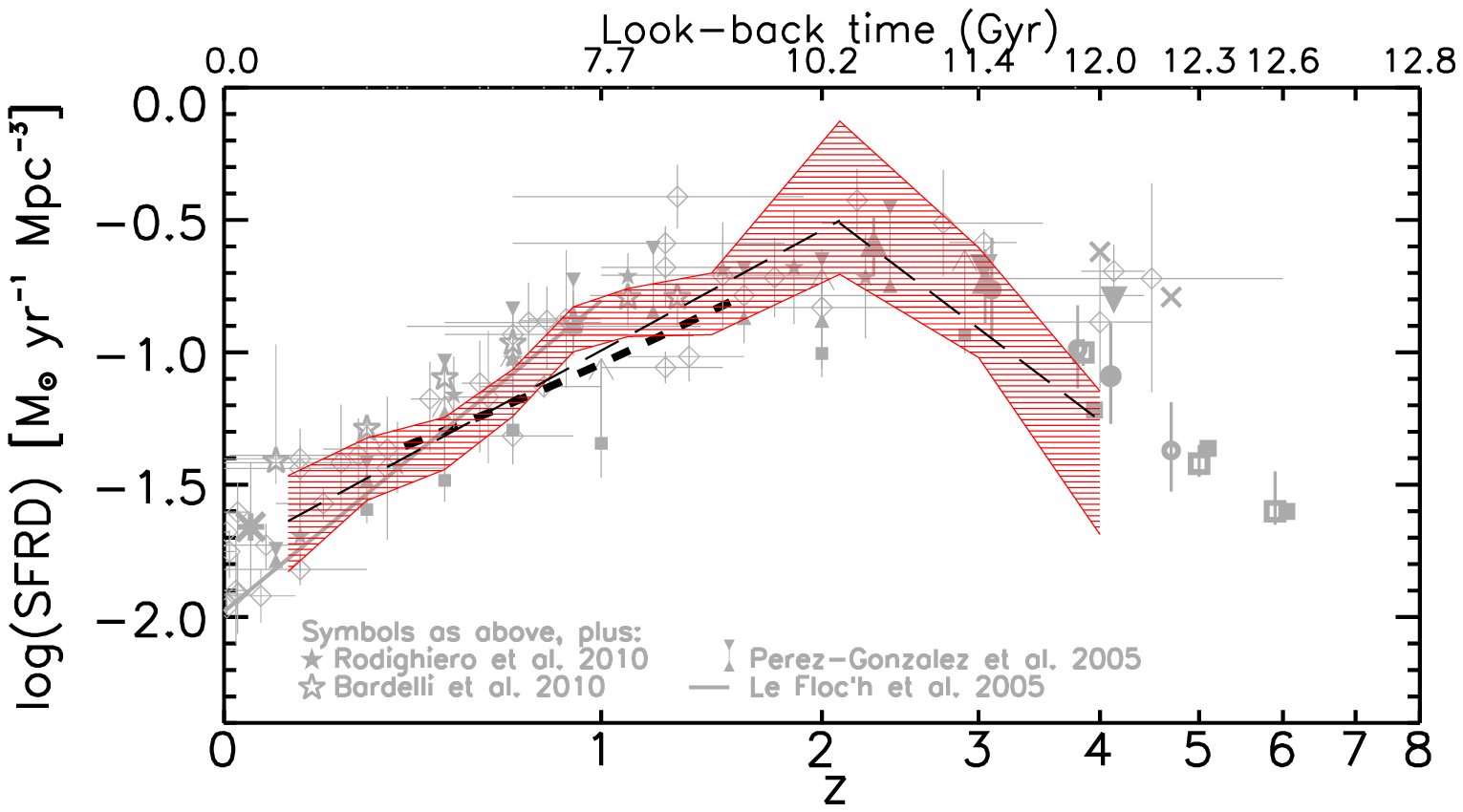}
\caption{{\it Top panel}. Total dust-corrected UV-derived SFRDs as a
  function of redshift from the VVDS Deep+Ultra-Deep sample (red
  filled circles, see Table~\ref{FUV_table_final}). Our uncertainties
  are explained in Sect.~\ref{lf}. The black dashed line is the
  SFRD(z) implied from the stellar mass density in
  \cite{ilbert2010_SM} (see text). We overplot other results from the
  literature, as detailed in the labels and at the end of the
  caption. Our SFRDs, as well as those from the literature, are
  derived using the FUV-band LDs converted into SFRD with the scaling
  relation from \cite{madau98}. All data have been homogenised with
  the same IMF \citep{salpeter1955}. {\it Bottom
    panel}.  Total dust-corrected UV-derived SFRDs as a
  function of redshift from the VVDS Deep+Ultra-Deep sample (red
  shaded area, corresponding to red circles and error bars in the top
  panel). The two long-dashed lines represent two fits to our SFRDs in
  the form $\propto(1+z)^{\beta}$ (see text for details). Gray points
  are from literature, derived from various SFR calibrators (UV,
  emission lines, IR, radio): they include the literature points as in
  the top panel, plus other works as in the labels. {\it List of
  references, top panel.} Compilation in \cite{hopkins2004} taking
  only the FUV and NUV determinations (grey open diamonds);
  \cite{steidel99} (LBG sample, cyan solid upside-down triangles);
  \cite{ouchi2004} (SDF and SXDF LBGs sample, thick grey crosses);
  \cite{wyder2005} (GALEX-2dFGRS, bold asterisk);
  \cite{schiminovich2005} (GALEX-VVDS at $z<1.2$ and HDF above, small
  solid squares with arrows); \cite{bouwens09_2z6} and
  \cite{bouwens11_beta4z7} (LBGs, orange open and filled squares,
  respectively); \cite{reddy_steidel2009} (LBG, solid magenta
  triangles; \cite{vanderburg10_LF_CFHTLS} (CFHTLS, black open
  circles); \cite{castellano11_betaz4} (LBG, blue open
  circles). {\it List of references, bottom panel.} All the points in
  the top panel, plus: grey thin diamonds comes from the entire
  compilation reported in \cite{hopkins2004};
  \cite{perez_gonzalez2005} (12$\mu$m, upside-down and normal solid
  triangles are the lower and upper limits; \cite{rodighiero2010} (LIR
  Spitzer VVDS-SWIRE \& GOODS, solid stars); \cite{bardelli2010}
  (radio VLA-zCOSMOS, open stars); \cite{leFloch05_IRLF}
  (24$\mu$m Spitzer CDFS, solid line; from their Fig.~14).}
\label{FUV_SFRDcorr_plot}
\end{figure*}

To transform FUV fluxes into star formation rates, we use the SFR
calibration of \cite{madau98}. It yields: 

\begin{equation} \displaystyle 
{\rm SFRD (z)}  = 1.4\, 10^{-28}\, {\rm
  LD_{FUV}(z)}\, 10^{0.4\ A_{FUV}(z)} , 
\end{equation}

\noindent where the SFRD is in $M_{\odot}$yr$^{-1}$Mpc$^{-3}$ units
and the LD in erg s$^{-1}$ Hz$^{-1}$ Mpc$^{-3}$. This formula includes
the dust attenuation $A_{FUV}(z)$, and assumes a \cite{salpeter1955}
initial mass function (IMF) including stars from 0.1 to 125 solar
masses.  The resulting dust-corrected SFRDs are shown in
Fig.~\ref{FUV_SFRDcorr_plot}. The values of uncorrected (i.e.,
$A_{FUV}(z)=0$) and dust-corrected SFRDs are given in
Table~\ref{FUV_table_final}.  For the SFRDs, we assume the same
uncertainties as the LDs (see Sect.~\ref{lf}), that is, we do not
include any dust reddening errors since it is the choice of the method
which dominates (see Sect.~\ref{dust_attenuation}). Still, we include
STY fit errors, Poisson noise, cosmic variance and weighting scheme
errors. We remind that the STY fit errors, when $\alpha$ is fixed,
include the span in LD (and so in SFRD) that one would have setting
$\alpha$ at the two extreme values ($-1.1$ and $-1.73$) discussed in
Sec.~\ref{lf}.  We overplot in the top panel of
Fig.~\ref{FUV_SFRDcorr_plot} dust-corrected UV-derived SFRDs from
other published works, assuming a \cite{salpeter1955} IMF. We refer to
\cite{hopkins2006} for a detailed comparison among different IMFs.

Like for the LD, we find a peak in the SFRD evolution at
$z\sim2$. This peak is preceded at earlier cosmic epochs by a rapid
increase of a factor 6 from $z\sim4.5$, then followed by a decrease by
a factor of 12 to $z\sim0$. We note that the VVDS Deep+Ultra-Deep SFRDs
are globally, within error bars, in agreement with the literature of
UV data sets obtained at various redshifts, but it is the first time
that the cosmic SFRD history is continuously traced from $z=0.05$ to
$z=4.5$ in a homogeneous way within one sample. We can therefore
assess the presence of a peak in the SFRD evolution at $z\sim2$,
without speculating about the different selection functions of
disparate surveys\footnote{Note that also uncertainties are often
computed in different ways in each galaxy survey, and they are
sometimes underestimated because not all-inclusive of many sources of
errors. This further increases the non-homogeneity of the SFRDs
determinations.}. We fitted the SFRD as a function of $z$ below and
above $z=2$, using SFRD$\propto(1+z)^{\beta}$. We find
$\beta=2.6\pm0.4$ at $z<2$, and $\beta=-3.6\pm1.9$ at $z>2$. These two
fits are shown in the bottom panel of Fig.~\ref{FUV_SFRDcorr_plot}.

We remark that this peak is the result of the combination of the LD
evolution and the dust attenuation evolution, but it is not uniquely
driven by the dust correction. First of all, we already found a clear
peak at $z\sim2$ in the evolution of the LD. Secondly, the peak of the
dust attenuation evolution does not coincide, in terms of time, with
the SFRD peak, because we find it at $z\sim1$ and not at $z\sim2$.
Surely, the shape of the dust attenuation evolution (steeply
decreasing from $z\sim2$ to $z\sim4$) enhances the relative decrease
of the SFRD with respect to the decrease of the LD in the same
redshift range, but it is not creating the SFRD peak by itself.

Although the scatter of measurements beyond $z\sim1$ is increasingly
larger, at $z\gtrsim3$ our measurements are lower than most other
measurements. This is mainly due to our use, at $z\sim3$, of a flatter
faint-end slope in the LF than other literature studies, and to a very
low $\phi^{*}$ at $z\sim4$ (see Sec.~\ref{LF_z2_discussion} for a
discussion about $\phi^{*}$ at $z>2$). The slope of the LF remains a
major source of uncertainty at these redshifts, but also the
computation of the dust attenuation. For example,
\cite{bouwens11_beta4z7} recently revised their previous $A_{FUV}$
computation \citep{bouwens09_2z6}: at $z\sim4$, their dust attenuation
correction factor (computed for $L\gtrsim0$) is now $\sim40$\% lower
than before. This leads to a lower total dust-corrected SFRD by $\sim0.2$
dex, giving further support to our result of a low SFRD at this redshift (see also
\citealp{castellano11_betaz4}). Interestingly, the dust attenuation at
$z\sim4$ in \cite{bouwens09_2z6} was closer to our $A_{FUV}$, but their
new determination of SFRD is closer to our SFRD value.

In the bottom panel of Fig.~\ref{FUV_SFRDcorr_plot}, we add also SFRD
data points in the literature, derived using other SFR calibrators
than the UV one, mainly at $z<2$. The non FUV-derived SFRD estimates
do not increase the scatter observed with the FUV-derived ones.  At
$z\le 2$, the cosmic SFRD evolution can be also derived from the stellar
mass density assembly, that is from the integrated history of the SFRD
down to a given epoch.  For instance, we take results from
\cite{ilbert2010_SM} (their Tables~2 and 3), who have computed the
stellar mass functions and $\rho_{star}$ densities over the
multi-wavelength COSMOS field \citep{scoville2007_COSMOS}.  From
$z=0.2$ to $z=2$, in each redshift bin we obtain $\rho_{star}$
densities in summing the individual stellar mass densities determined
for the ``quiescent, ``intermediate activity'' and ``high activity''
galaxy populations. Assuming a linear fit, we recover its evolution
with redshift, $\rho_{star}(z)$, and the implied SFRD(z) with the
recipe detailed in \cite{wilkins2008_SM}\footnote{Stellar masses in
\cite{ilbert2010_SM} are computed using a \cite{chabrier2003_IMF}
Initial Mass Function, that we convert to a \cite{salpeter1955} IMF.}.
Note that we assume that 30\% of the created stellar mass returned
into the interstellar medium: as suggested by
\cite{prantzos2008_review}, this is the average fraction derived using
different IMF. As shown in Fig.~\ref{FUV_SFRDcorr_plot}, the resulting
inferred SFRD(z) agrees on average with our direct SFRD(z)
estimate. Nevertheless, there is some discrepancy ($\sim0.2$ dex)
around $z\sim1$, which happens at the cosmic dust attenuation peak
(see Fig.~\ref{AFUV_plot}) within the star-forming population.  

In summary, the VVDS Deep+Ultra-Deep survey traces continuously the
cosmic dust-corrected SFRD over $\sim$12~Gyr in a comprehensive way,
with homogeneous treatment of data, sources of errors, and dust
attenuation correction. A clear peak of the SFRD emerges at $z\sim2$,
preceded at earlier cosmic times by a rapid increase by a factor
$\sim$6 within $\sim$2~Gyr, and followed by a decrease by a factor
$\sim$12 within $\sim$10~Gyr, down to $z\sim0$. For the first time,
the SFRD evolution is genuinely established over 12 Gyr.

\subsection{The contribution of galaxies with different 
luminosities}\label{SFRD_lum}

The relative contributions to the total SFRD of distinct galaxy
populations vary with redshift, and it is their combination that
shapes the global SFRD evolution.  \cite{tresse2007} have assessed the
contributions of galaxies with different luminosities to the non
dust-corrected FUV-band LDs, showing, for instance, that at
$z\sim3.5$, the different FUV populations equally contribute to the
total LD. Then, as the Universe is ageing, the contribution of the
luminous FUV population ($M_{FUV}<-19$) dimes, while the one of the
fainter population increases. With our present data, we can directly
study the contribution of galaxies with different luminosities to the
dust-corrected SFRD. We defer to a future work the analysis on how
galaxies contribute to the total SFRD as a function of their stellar
mass.

\begin{figure} \centering
\includegraphics[width=\linewidth]{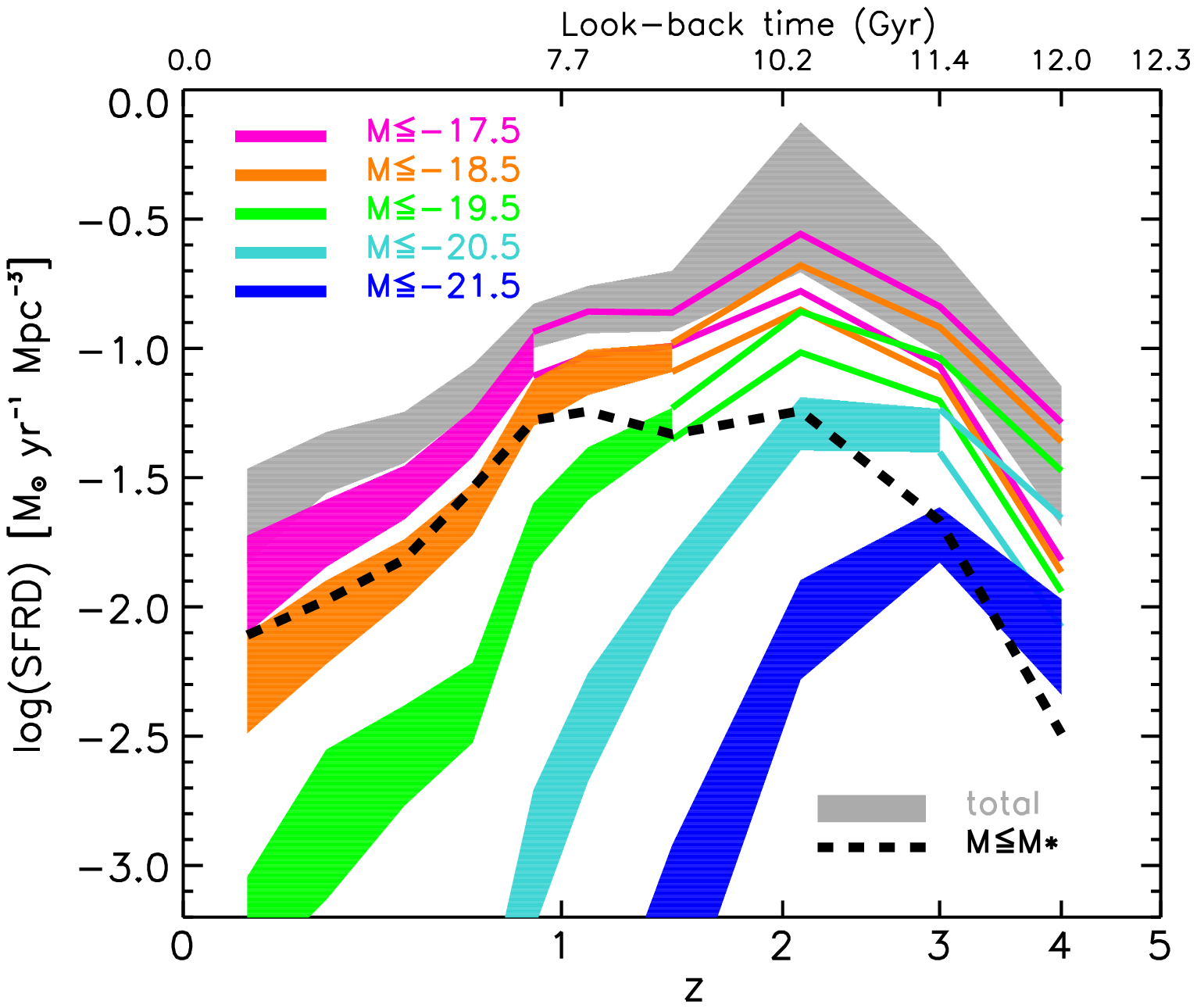}
\caption{The dust-corrected SFRD from galaxies with different
luminosity limits, as described in the labels. The background gray
shaded area is the total SFRD, while the SFRD for different luminosity
cuts decreases for brighter galaxies. The luminosity limits are kept
fixed at all redshifts. The shaded area for each SFRD accounts for the
same kind of errors as for the total SFRD (STY ellipses, cosmic
variance, Poisson noise, weighting scheme, and at $z>1.7$ also the
span of $\alpha$ between $\alpha=-1.1$ and $\alpha=-1.73$). When
the luminosity limit is fainter than the LF magnitude bias (see
Sec.~\ref{lf_methods}) at a given redshift, the area between the lower
and upper 1-$\sigma$ error is not filled but left empty. The
dashed black line is the SFRD from galaxies brighter than $M^{*}$ at
each redshift (i.e., the luminosity cut varies with redshift following
$M^{*}$). At each redshift, error bars on this SFRD are similar to
those for the cut in luminosity closest to $M^{*}$, and we do not
overplot them for clarity. We apply to all the points the same dust
correction as for the total SFRD, as we do not detect a clear
dependence of E(B-V) on FUV luminosity (see discussion in
Sec.~\ref{dust_attenuation}). } \label{FUV_SFRDcorr_plot_contr}
\end{figure}

Fig.~\ref{FUV_SFRDcorr_plot_contr} shows the dust-corrected SFRD of
galaxies with different luminosity upper limits: $M_{FUV} \leq -17.5,
-18.5, -19.5, -20.5, -21.5$ and $M_{FUV} \leq M^{*}_{FUV}$ (we remind
that $M^{*}_{FUV}$ varies with $z$, see Table~\ref{FUV_table_final}).
We apply to each SFRD the same dust correction derived for the total
SFRD, since we do not detect any significant $E(B-V)$ colour
dependence with the FUV luminosity (see Sect.~\ref{dust_attenuation}).
We see that, as times goes by, the total SFRD is dominated by fainter
and fainter galaxies. In fact, bright galaxies with $M_{FUV} \le
-21.5$ contribute little to the total SFRD for $z\lesssim2$, and this
holds for galaxies with $M_{FUV} \le -20.5$ at $z\lesssim1$ and with
$M_{FUV} \le -19.5$ at $z\lesssim0.8$.

\begin{figure} \centering
\includegraphics[width=\linewidth]{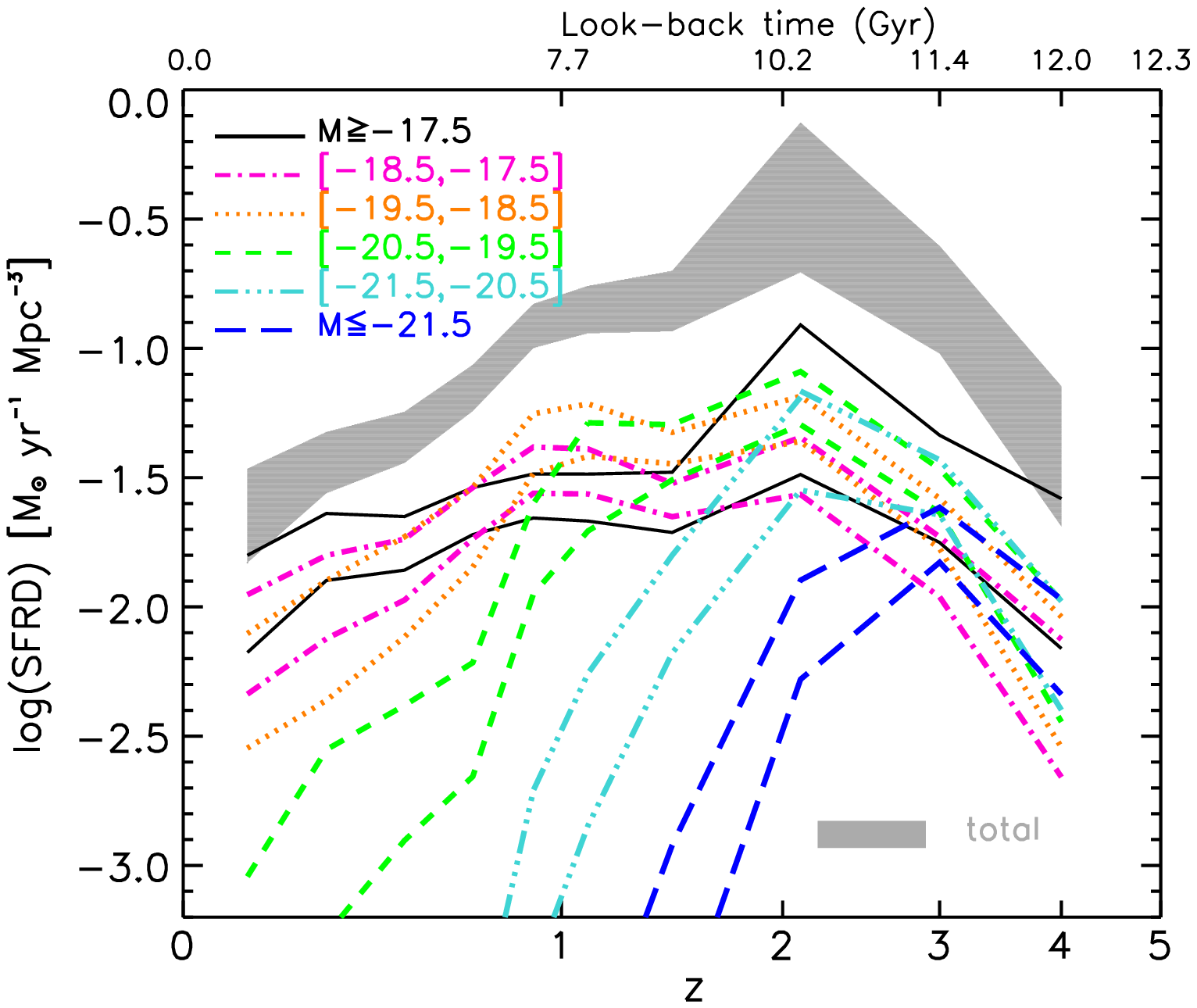}
\caption{As in Fig.~\ref{FUV_SFRDcorr_plot_contr},
but this time galaxies are divided in narrow luminosity bins, as
indicated in the labels. The first and last bins are actually the
extremes of the luminosity distribution (the faintest and the
brightest galaxies). Also in this case, as in
Fig.~\ref{FUV_SFRDcorr_plot_contr}, the errors include all sources of
uncertainty. The two lines for each luminosity bin define the region
of $\pm 1$-$\sigma$ error. Not to crowd the plot, as the plotted
quantities have larger error bars than in
Fig.~\ref{FUV_SFRDcorr_plot_contr}, here we do not distinguish when
the luminosity bin considered is fainter or brighter than the LF
bias. This can be inferred by Fig.~\ref{FUV_SFRDcorr_plot_contr}. } 
\label{FUV_SFRDcorr_plot_perc} 
\end{figure}

To investigate further these trends, we made 6 classes of galaxies
within the following $M_{FUV}$ luminosity ranges: $\geq -17.5$,
$[-17.5,-18.5]$, $[-18.5,-19.5]$, $[-19.5,-20.5]$, $[-20.5,-21.5]$ and
$\leq -21.5$ (see Fig.~\ref{FUV_SFRDcorr_plot_perc}).  We describe
below the three main cosmic eras of the SFRD history.

{\it From $z=4$ to $z=2$}. During this period, the SFRD from each
galaxy subclass increases, with the only exception of the brightest
galaxies ($M_{FUV}<-21.5$), whose SFRD starts decreasing at $z\sim3$.
Even including all sources of uncertainties, the contribution of these
brightest galaxies to the total SFRD is a maximum of 20\% at $z\sim3$
and decreases to a maximum of 10\% at $z=2$. The net effect is that
the total SFRD increases from $z=4$ to $z=2$, because the SFRDs of the
main contributors ($M_{FUV}>-21.5$) increase as well.

{\it From $z=2$ to $z=1$}. The period corresponds to the change of the
shape and normalisation of the FUV LFs (see Figs.~\ref{FUV_LF_plot}
and \ref{SFRD_picture}). Thus, the trends are more differentiated from
class to class: the SFRDs for [$-19.5$,$-17.5$] stay about constant, for
$>-17.5$ and [$-20.5$,$-19.5$] mildly decrease, for $<-20.5$ sharply
drop. The net effect is however a global decrease, the highest
star-forming populations being shut down (galaxies brighter than
$M_{FUV}=-20.5$ contribute maximum by 3\%).

{\it From $z=1$ to $z\sim0$}. Through this period, the individual
SFRDs reach a general decreasing trend, particularly steep for [$-20.5$,
$-19.5$] galaxies. As times goes by, the less galaxies are forming
stars, the larger is their contribution to the total SFRD. The net
effect is that the total SFRD decreases from $z=1$ to $z\sim0$.

To summarise, the total SFRD evolution (with an increase from $z=4$ to
$z=2$, a peak at $z=2$ and a decrease from $z=2$ to $z=0$) is driven
by different galaxy classes at different epochs. In particular, the
percentage contribution to the total SFRD of all galaxy classes is
roughly the same at $z=2$, but the peak of the total SFRD at this
redshift is mainly due to a similar peak of the SFRD from galaxies
with $-21.5 \leq M_{FUV}<-19.5$ ($L\gtrsim L^{*}_{z=2}$). In contrast,
the contribution of the most luminous star-forming galaxies reaches
its maximum at higher redshift ($z\sim3$). We note that the faintest
galaxies ($M_{FUV}>-17.5$) are the only ones that show a continuously
decreasing SFRD from $z=2$ to $z=0$.

Fig.~\ref{FUV_SFRDcorr_plot_perc} shows that at $z>3$ each FUV
luminosity class shows a very similar SFRD trend with $z$, while for
$z<3$ the contributions of the various classes to the SFRD become more
and more dissimilar as times goes by. This implies i) the presence of
a very high-$z$ population not yet strongly affected by star formation
regulation/quenching mechanisms and ii) a downsizing pattern in the
SFR history.

We remark that the contribution of the faintest galaxies
($M_{FUV}>-17.5$) is clearly affected by the shape (and its evolution)
of the LF. We refer the reader to \cite{tresse2007} for an extensive
analysis of the role of $\alpha$ in the relative contribution of faint
galaxies to the total LD. The difficulty in constraining $\alpha$, and
the spread in typical luminosity of the targeted populations, are
surely two important sources of uncertainty in the literature in the
measurement of the total SFRD shape in this redshift range.

\section{Summary \& Discussion}\label{discussion}

In this work, we have computed the rest-frame FUV luminosity functions,
luminosity densities and dust-corrected star formation rate densities
within a single and deep redshift survey, merging the VVDS Deep and
VVDS Ultra-Deep data sets (with overall $17.5\leq I_{AB} \leq 24.75$),
over a 12~Gyr cosmic time baseline ($0.05<z\leq 4.5$). We also derived
the average dust attenuation in the entire redshift range explored. Our data
constitute a large improvement with respect to those used in previous
VVDS studies: i) the Deep survey includes more redshifts, ii) repeated
observations (leading to 100\% secure redshift measurement for
previous lower quality spectra) enabled us to correct our $n(z)$ for
flags 1 and 2, iii) we use the Ultra-Deep sample, pushing 0.75~mag deeper, iv) we have
deeper optical photometry, in particular in the $u^{*}$-band and
$z$-band, and v) we have used new near-infrared deeper photometry.

\begin{figure} \centering
\includegraphics[width=\linewidth]{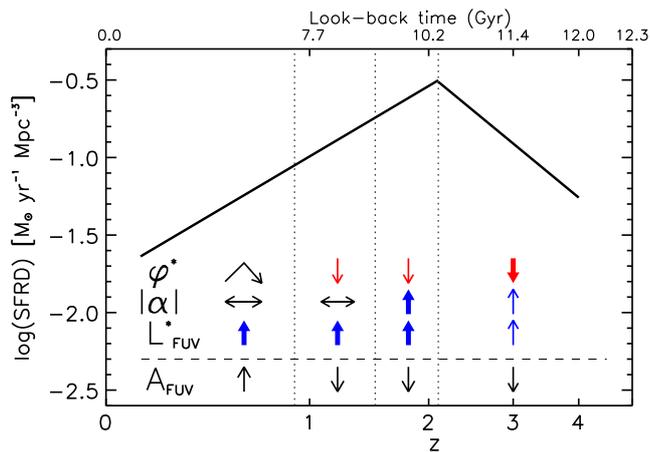}
\caption{Scheme of the total dust-corrected SFRD as a
function of redshift, as found in this work (solid broken line, see
Sec.~\ref{SFRD_dust_sec}). We distinguish four epochs, separated with vertical dotted
lines. At each epoch, arrows indicate the behaviour of the three
Schechter parameters and $A_{FUV}$: arrows pointing up mean that the
parameter is increasing with increasing redshift, arrows pointing down
mean that the parameter is decreasing. Red arrows indicate trends that
in principle should decrease the total SFRD, the opposite is for blue
arrows. Thick arrows indicate which parameter is driving the SFRD
evolution at that given epoch. In this raw scheme, we neither consider
$A_{FUV}$ dominating, nor making increase or decrease the SFRD: clearly, its
trend enhances or weakens the behaviour of the LD when it is
transformed to dust-corrected SFRD, but the peak at $z\sim2$ is
already found in the LD. } 
\label{SFRD_picture} 
\end{figure}

Our  results are summarised as follows. 
\begin{itemize}

\item[-] We find a flat and constant faint-end slope in the FUV-band LF
      at $z<1.7$ ($\alpha\sim1$). We verified that this is unlikely to
      be the result of missing faint galaxies from our $I$-band
      selection, and that dust may have a role (see discussion in
      Sec.~\ref{LF_discussion}). At the same time, $M^{*}_{FUV}$
      increases by $\sim 1.5$ mag from $z\sim0$ to $z\sim1.2$, while
      $\phi^{*}$ starts decreasing at $z>0.7$ (more than a factor of 2
      decrease from $z\sim0.7$ to $z\sim1.5$). 

\item[-] At $z>1.7$, we set $\alpha$ evolving with $(1+z)$, and this
      way it becomes as steep as $-1.73$ at $z\simeq4$, consistent
      with values from deep photometric studies. In the meanwhile,
      $M^{*}_{FUV}$ keeps on brightening (by an added 2.5 mag up to
      $z\sim4$), and $\phi^{*}$ decreasing (by another factor of
      $\sim40$ up to $z\sim4$). We find that at $z=2,3,4$ our
      $M^{*}_{FUV}$ is on average brighter than what has been found in
      previous works, and that $\phi^{*}$ is on average smaller, in
      particular at $z\sim4$. A summary raw scheme of these trends
      within $0<z<4$ is shown in Fig.~\ref{SFRD_picture}.

\item[-] We find that at $z\sim3$, while the projected number counts
     of I-selected galaxies like VVDS are at least twice larger than
     the projected number counts of LBG-selected galaxies (see
     Le~F\`evre et al., in prep.), the VVDS LF has about 50\%
     higher density at $M_{FUV}\sim-22$ than the LF derived from LBG
     counts. To interpret this apparent discrepancy, we verified the
     importance of the redshift
      distribution $n(z)$ shape when transforming number
      counts (as a function of the observed magnitude) into a LF, or
      vice-versa. Also, when comparing number counts of two different
      samples in a given redshift range, one has to take into account
      the $n(z)$ of the two data sets.

\item[-] We derived the evolution of the dust-attenuation in the FUV-band
      ($A_{FUV}$) in the range $0.05<z\leq 4.5$, using a SED fitting
      method, in a consistent way from a single survey with a well
      controlled selection function. We find a continuous increase of $A_{FUV}$ by 
      $\sim1$ mag from $z\sim4.5$ to $z\sim1$ (with the increase
      being very slow at $1.5>z>1$), then a decrease by the same
      amount from $z\sim1$ to $z\sim0$. This is the first time that
      the $A_{FUV}$ evolution has been assessed homogeneously on such
      a broad redshift range.  
	
\item[-] We traced the dust-corrected SFRD evolution over the past
      $\sim12$~Gyr. It can be schematically fitted as SFRD(z) $\propto
      (1+z)^{\beta}$, with $\beta=2.6\pm0.4$ at $z<2$ and
      $\beta=-3.6\pm1.9$ at $z>2$ (see Fig.~\ref{FUV_SFRDcorr_plot}
      and \ref{SFRD_picture}). Thanks to the homogeneity of our data
      over such a cosmic time, we have been able to unveil the presence
      of a peak at $z\sim2$ in the cosmic SFR history. This peak is
      preceded by a rapid increase by a factor 6 from $z\sim4.5$, then
      followed by a general decrease by a factor 12 to $z\sim0$. We
      remark that the epoch of the peak of the dust-corrected SFRD
      ($z\sim2$) does not coincide with that of the maximum of the
      dust attenuation evolution ($z\sim1$), and that a peak at
      $z\sim2$ is already present in the evolution of our LD.

\item[-] Studying the contribution to the total SFRD of galaxy
      populations with different properties, we find that as times
      goes by, the total SFRD is dominated by fainter and fainter
      galaxies. Moreover, the presence of a SFRD peak at $z\sim2$ is
      mainly due to a similar peak within the population of galaxies with
      $-21.5 \leq M_{FUV} \leq -19.5$, while the
      most extreme star-forming galaxies reaches their maximum
      activity at higher redshift.

\end{itemize}

Our data therefore consistently show a peak in the LD and SFRD at
$z\sim2$, where the LF is well constrained. While the increase in SFRD
from $z=4$ to $z=2$ is not in question, the exact amplitude of this
increase remains to be investigated, as the faint end slope of the LF
is still unconstrained even from our very deep spectroscopic
survey. Other measurements of the SFRD beyond $z=2$, using
e.g. photometric redshifts derived from multi-band imaging (for
example \citealp{gabasch2004_LF,sawicky2006_LF,bouwens09_2z6}), go
deeper than spectroscopic samples but carry significant associated
uncertainties either on the photometric redshifts or because of the
small size of the fields and/or samples. It is then fair to say that
the exact increase of the SFRD from higher redshift to $z\sim2$
remains to be robustly quantified.

The correct determination of the shape of the SFRD evolution is
necessary to understand which physical processes mostly affect galaxy
evolution. The SFRD is the result of the transformation of gas into
stars and therefore requires a significant gas reservoir to sustain a
strong star formation rate for a long time. A number of processes are
expected to modify the gas reservoir hence the SFR, including the
efficiency of star formation, cold accretion along the cosmic web
filaments, mergers with gas-rich galaxies, stellar feedback, SN
feedback blowing gas out from the galaxy core, AGN feedback, cosmic
photoionizing radiation, or environment effects which may result in
star formation quenching (e.g., among many others,
\citealp{white_and_frenk91,efstathiou92_photo,cole1994,cole00,dimatteo05_agnwinds,baugh2006,cox08_merger,dekel2009_nat,deravel2009}). The
exact balance of these different processes along cosmic time will
result in the observed SFRD. The SFRD peak that we find at $z\simeq 2$
is produced by galaxies with $L$ close to or even brighter than
the $L^{*}$ of that redshift, requiring that significant gas
reservoirs still exist at this epoch and are probably replenished by
cold accretion and wet mergers, while feedback or quenching processes
are not yet strong enough to lower the SF. The knee of the
rest-frame FUV LF shape is smoothed away at $z\gtrsim2$, i.e. there is
not anymore a clear distinction between the high and intermediate
star-forming galaxy densities. It does not mean that the high-z
gas-rich galaxies form a homogeneous population, but likely that the
long time-scale ($t>3$Gyr) processes do not heavily affect the galaxy SFR
yet.  Knowing the SFRD, we may hope to identify the relative
contribution of these different processes at different epochs.

Using simulations (semi-analytical galaxy evolution models, smoothed
particle hydrodynamics simulation), several authors have attempted to
reproduce the observed Cosmic Star Formation History with theoretical
predictions (e.g.
\citealp{baugh2005_sfrdmod,somerville08_mod,fontanot09_downsizing,hopkins_2010_mod,vandevoort2011_sfrd_sim})
As a recent example, \cite{weinmann2011_sSFR} use semi-analytical
galaxy evolution models to predict the SFRD evolution (see their
Fig.~10). They start from a standard model of galaxy evolution, and
produce slightly different versions of it by tuning one or more
ingredients at a time (such as star formation efficiency, stellar
feedback, merger processes...).  Although the main goal of their work
is not to reproduce the SFRD evolution, from their Fig.~10 it is clear
that models in which different physical processes have been enhanced
or depressed predict significantly different Cosmic Star Formation
Histories. All of their models reproduce qualitatively the fast SFRD
increase from $z\sim0$ to $z\sim1$, but at $z>1$ the predicted SFRDs
have different behaviours, showing growths and decreases at different
rates, in some cases a plateau, in some cases a more or less
pronounced peak at different possible redshifts. It is worth noticing
that one of the models (where feedback has been tuned) predicts a
sharp peak in the SFRD evolution at $z\sim1.5$, qualitatively similar
to the one that we find at $z\sim2$.  It is not the aim of this paper
to compare in details our findings with model predictions, and we will
address this issue in more details in a future work. Here we want only
to stress the importance of bringing strong observational constraints
on the SFRD from a unique and homogeneous galaxy sample covering a
large cosmic time of $\sim12$ Gyr, which will need to be reproduced by
next generation models.

Another remarkable finding of our work is that the peak of the dust
and the SFR do not coincide, differently from what one could have
naively thought in the case the dust is immediately released into the
ISM a short time after supernova explosions of massive, short-lived
stars which dominate the SFR. These two peaks are separated by
$\sim2.5$~Gyr, which is a long period if one considers that the dust
production rate peak is below 1~Gyr for SNII, but it is at 3-4 Gyr for
intermediate-mass stars (see \citealp{dwek1998}).  Nevertheless the
dust reaches a sort of plateau from $z\sim1.5$ (with a maximum at
$z\sim1$), i.e. 1~Gyr after the peak of the SFRD. Recently,
\cite{fukugita2011} suggested that the dust must survive on much
longer time scales than what has been previously thought and that half
of the dust could be produced by SNII and the other half by
intermediate-mass (1-8 solar masses), long-lived stars.  If we assume
that the SNII dust production peaks very shortly after the SFRD peak,
then the dust peak that we observe at $z\sim1$ is likely due to
intermediate-mass, long-lived stars producing their peak of dust on a
delayed time. In particular, the AGB stars release dust through
intense mass loss, and most efficiently at the very end stages of
evolution \citep{gall2011rev}. The peak of attenuation is higher
than SNII simply because these stars are much more numerous than very
massive, short-lived OB stars, assuming a universal IMF.  It would
explain also that at $z>2$, the low level of dust attenuation is
mainly due to dust produced by SNII, while at $z<2$, it is resulting
from the combination of dust from SNII and intermediate-mass stars.
Our findings, combined with the above-mentioned times scales, may
imply that dust is not only quickly formed (thanks to SNII ejecta and
remnants), but also quickly destroyed, because the peak of attenuation
is only $\sim2.5$Gyr after the SFRD peak, a shorter time-scale than
the typical one for dust production by intermediate-mass stars (3-4
Gyr).  Surely, like in the case of the SFRD evolution, our findings
about the general evolution of the FUV-band dust attenuation in such a
broad cosmic epoch ($\sim12$ Gyr) will constitute an important
reference for future models.

\begin{acknowledgements}
  We would like to thank the referee for useful suggestions and
  comments, that improved the manuscript. We thank S.~Boissier for
  useful discussions. OC thanks V.~Buat, D.~Burgarella and L.~Cortese
  for interesting conversations. This work has been partially
  supported by the CNRS-INSU and its Programme National
  Cosmologie-Galaxies (France) and by INAF grant COFIN 2010.
\end{acknowledgements}

\bibliographystyle{aa}
\bibliography{18010bibl}

\appendix


\section{The weighting schemes}\label{appendix_weights}

We refer the reader to \cite{lefevre2005a} and Le~F\`evre et al. (in prep.) for
details on the quality flags assigned to redshift measurements in the
VVDS Deep and Ultra-Deep surveys, respectively. We detail below the
weighting schemes to be applied to our statistical analyses. They
account for the selection function of the photometric sources targeted
to acquire their spectrum (the Target Sampling Rate, TSR), and for the
success to measure a reliable redshift from the spectrum (the
Spectroscopic Success Rate, SSR).  They are derived for the Deep
survey, the Ultra-Deep survey, and for the merged Deep+Ultra-Deep
sample.

To derive the weighting schemes, we made use of the photometric
redshifts, $zphot$, computed as described in \cite{ilbert2006zphot},
but using the more recent T0005 release of CFHTLS photometric data
($u^{*}, g', r', i', z'$) and the latest near-infrared
photometric data available from WIRDS ($J$, $H$ and
$K_{s}$, \citealp{bielby2011_WIRDS}).

We emphasise that we correct the Deep and Ultra-Deep distributions of
photometric redshifts, $n(zphot)$, for their failure rate computing
the ratio between the spectroscopic redshifts measured with a $>97\%$
confidence level (flags 3, 4), $n(z_{f=3,4})$ and their corresponding
photometric redshifts, $n(zphot_{f=3,4})$.

\subsection{Weights for the Deep survey}
\label{appendix_weights_deep}

The Deep survey TSR$_{d}$ is defined as $N_{target}^{d}$/$N_{phot}^{d}$, where
$N_{phot}^{d}$ is the number of photometric sources in the Deep
survey with $17.5 \leq I_{AB} \leq 24.0$, over
$\sim$2200~arcmin$^{2}$ of sky area, and $N_{target}^{d}$ is the
number of photometric sources targeted for spectroscopic
observations. The TSR$_{d}$ depends on the projection of the angular
size of each object on the x-axis of the image (`x-radius', see also 
\citealp{ilbert2005}). The weight associated to TSR$_{d}$ is
$w_{TSR}^{d}(r)=1/TSR_{d}(r)$, where $r$ is the x-radius defined above.

The Deep survey SSR$_{d}$ is defined as $N_{spec}^{d}$/$N_{target}^{d}$, where
$N_{spec}^{d}$ is the number of targets with a reliable spectroscopic
redshift measurement in the Deep survey. The SSR$_{d}$ is a function
of both selection magnitude and redshift \citep[see][]{ilbert2005}. As
described in Le~F\`evre et al. (in prep.), a fraction of low confidence level
spectroscopic redshifts (flags 0, 1, 2) at $z\ge1.4$ from the Deep
survey have been observed again, leading to a 100\% confidence level
in the redshift measurement.  Comparing the old redshift distribution 
$n(z_{f=1,2}^{reobserved})$ to the new one for this subsample, we have
remodulated the full $n(z_{f=1,2})$ partially using the $n(zphot)$
for $z<1.4$ sources.  We have also remodulated the single emission-line 
redshifts (flags 9) $n(z_{f=9})$ according to its
$n(zphot_{f=9})$.  This gives:
 
\begin{eqnarray} \displaystyle 
N_{spec}^{d}(m,z) & = & N_{f=1,2}(m,z) \times  n(m,z_{f=1,2}^{reobserved}) / n(m,z_{f=1,2}) +{}  \nonumber \\
   &  &  {}+ N_{f=9}(m,z)\ \times \ n(m,zphot_{f=9})/ n(m,z_{f=9}) + {} \nonumber\\
   &  &  {}+ N_{f=3,4}(m,z). \nonumber
\end{eqnarray}
The weight associated to SSR$_{d}$ is $w_{SSR}^{d}(m,z)=1/SSR_{d}(m,z)$.

In summary, for the Deep survey, we have $N_{spec}^{d} \subset
N_{target}^{d} \subset N_{phot}^{d}$ and we apply the weight $w_{d}(r,m,z) =
w_{TSR}^{d}(r)\ w_{SSR}^{d}(m,z)$ to each galaxy in $N_{spec}^{d}$.

\subsection{Weights for the Ultra-Deep survey}
\label{appendix_weights_udeep}

The Ultra-Deep survey covers a 576~arcmin$^{2}$ sky area embedded in
the Deep survey area. It is purely flux limited at $23.00 \leq
i'_{AB} \leq 24.75$ with $N_{phot}^{ud}$ photometric sources. 
Several spectroscopic sources with $23 \leq I_{AB} \leq 24$ had  already been observed 
in the VVDS Deep survey, when we started the
Ultra-Deep observations. By excluding them, we obtained a reduced
catalogue with $N_{phot}^{parent}$ sources available for the
Ultra-Deep spectroscopic target selection.
The Ultra-Deep Photometric Sampling Rate (PSR$_{ud}$) is defined as 
$N_{phot}^{parent}/N_{phot}^{ud}$. The PSR$_{ud}$ depends on the
$I$-band apparent magnitude. The weight associated to PSR$_{ud}$ is
$w_{PSR}^{ud}(m)=1/PSR_{ud}(m)$.

The Ultra-Deep survey TSR$_{ud}$ is defined as 
$N_{target}^{ud}$/$N_{phot}^{parent}$, where $N_{target}^{ud}$ is the
number of photometric sources targeted for Ultra-Deep spectroscopic
observations. The TSR$_{ud}$ does not depend on any parameter, it is a
constant value (6.5\%).  The weight associated to TSR$_{ud}$ is
$w_{TSR}^{ud}=1/TSR_{ud}$.

The Ultra-Deep SSR$_{ud}$ is defined as $N_{spec}^{ud}$/$N_{target}^{ud}$, where
$N_{spec}^{ud}$ is the number of targets with a reliable spectroscopic
redshift measurement in the Ultra-Deep survey.  Here $N_{spec}=
N_{f=1.5,2,3,4,9}$, where flags~1.5 is a flag~1 with a photometric
redshift in agreement with the spectroscopic redshift
(Le~F\`evre et al., in prep.). The SSR$_{u}$ is a function
of both selection magnitude and redshift. The associated weight is
$w_{SSR}^{ud}(m,z)=1/SSR_{ud}(m,z)$.

In summary, for the Ultra-Deep survey, we have $N_{spec}^{ud} \subset
N_{target}^{ud} \subset N_{phot}^{parent} \subset N_{phot}^{ud}$ and
we apply the weight $w_{ud}(m,z)=w_{PSR}^{ud}(m)\ w_{TSR}^{ud}\ w_{SSR}^{ud}(m,z)$
to each galaxy in $N_{spec}^{ud}$.

\begin{figure}
\centering
\includegraphics[width=\linewidth]{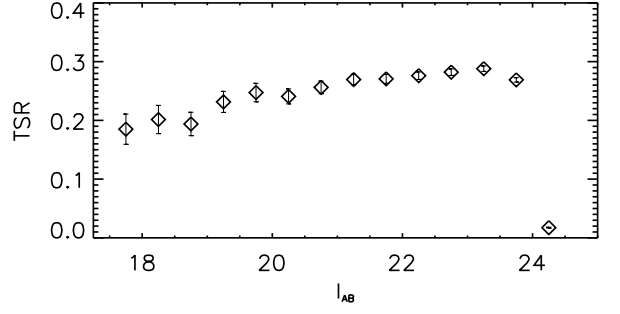}
\caption{Target Sampling Rate (TSR) for the merged Deep+Ultra-Deep
  catalogue as a function of the $I$-band apparent magnitude.  Given
  the different flux limit of the two surveys (see text), this TSR is
  obtained using Deep data at $I_{AB}<23$, both Deep and Ultra-Deep
  data at $23 \leq I_{AB}\leq 24$, and Ultra-Deep data at $ I_{AB}>
  24$. The last point is low because the Ultra-Deep survey TSR is
  much lower than the Deep survey TSR. }
\label{TSR_fig}
\end{figure}

\subsection{Weights for the combined Deep and Ultra-Deep surveys}
\label{appendix_weights_merged}

We combine  the two surveys and  we derive an adapted weighting scheme
to take advantage of both (i) the large magnitude range covered by the
Deep  survey ($17.5\le I_{AB}  \le 24.0$)  over  a large sky area, and
(ii) the depth reached by the Ultra-Deep  survey (0.75 mag deeper than
the Deep) over a smaller embedded sky area.

As a first step, we need to merge the two catalogues of galaxies in
accounting for their different depth, especially for their different
LF bias limit (see Sect.~\ref{lf_methods}). Since the latter depends
on the studied redshift bin, we merge the catalogues in each redshift
bin in which we explore the LF, as follows.  For the Ultra-Deep
survey, we compute the LF bias limits with the method described in
Sect.~\ref{lf_methods}.  For the Deep survey, we adopted more
conservative limits with respect to those obtained with the above
mentioned method, because of the magnitude range $23 \leq I_{AB}\leq
24$ common with the Ultra-Deep survey. In the Deep and Ultra-Deep
samples we keep only galaxies brighter than the respective bias limit,
and we merge the two sub-samples into a single catalogue.  From now
on, the merged catalogue is considered with flux limits within $17.50
\leq I_{AB} \leq 24.75$, over the Deep survey effective sky area.

As a second step, we need to slightly modify the TSR weighting scheme
for the merged catalogue since the Ultra-Deep area is embedded in the
Deep area.  The Deep+Ultra-Deep TSR$_{d+ud}$ is $(N_{target}^{d} +
N_{target}^{ud}) / N_{phot}^{full}$, where $N_{phot}^{full}$ is the
full photometric catalogue at $17.50 \leq I_{AB} \leq 24.75$ over
$\sim2200$ arcmin$^2$.  The TSR$_{u+ud}$ depends on the $I$-band
apparent magnitude (see Fig.~\ref{TSR_fig}). At $17.5\leq
I_{AB}<23.0$, it corresponds to TSR$_{d}$ when we compute it as a
function of magnitude instead of angular size, while at
$24.00<I_{AB}\leq24.75$, it corresponds to the constant TSR$_{ud}$.
At $23 \leq I_{AB} \leq 24$, it corresponds to the combination of both
TSRs, i.e. at the numerator we have all targets of both the Deep and
Ultra-Deep surveys.  The weight associated to TSR$_{d+ud}$ is
$w_{TSR}^{d+ud}(m)=1/TSR_{d+ud}(m)$.  We applied this weight according to
the $I$-band apparent magnitude to all Deep survey galaxies and to
Ultra-Deep galaxies which are brighter than the LF bias limits of the
Deep survey.  To the Ultra-Deep galaxies which are fainter than the
Deep LF bias limits we applied the $w_{TSR}^{u+ud}$ shown for $I_{AB}
\geq 24$.

In summary, for the merged Deep+Ultra-Deep catalogue, we apply the
weight $w_{d}^{merged}(m,z) = w_{TSR}^{d+ud}(m)\ w_{SSR}^{d}(m,z)$ to
$N_{spec}^{d}$ and the weight $w_{ud}^{merged}(m,z) = w_{TSR}^{d+ud}(m)\
w_{SSR}^{ud}(m,z)$ to $N_{spec}^{ud}$.

\begin{figure} 
\centering
\includegraphics[width=\linewidth]{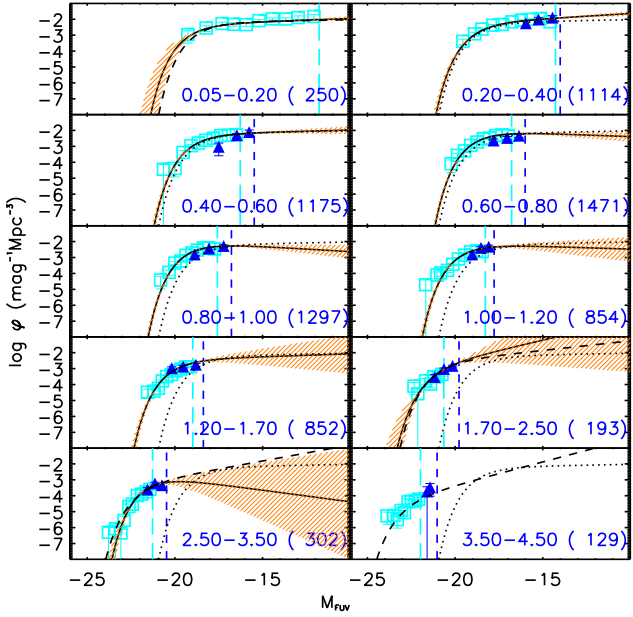}
\caption{Rest-frame FUV $1/V_{max}$ LF data points for the Deep survey
  (cyan empty squares) and the Ultra-Deep survey (blue filled
  triangles), from $z=0.05$ to $z=4.50$. There is no Ultra-Deep data
  in the lowest redshift bin. In each panel, the FUV absolute
  magnitude LF bias limit (see Sect.~\ref{lf_methods}) is shown as a
  vertical cyan long-dashed line for the Deep survey and as a blue
  short-dashed line for the Ultra-Deep survey.  The number quoted in
  parenthesis is the total number of used galaxies.  We overplot the STY
  determinations of the merged Deep+Ultra-Deep sample, as illustrated
  in Fig.~\ref{FUV_LF_plot} and tabulated in
  Table~\ref{FUVNUV_LF_table}): (i) with free Schechter parameters
  (solid lines) and its associated error (shaded area), (ii) in
  setting the faint-end slope at $z>1.7$ (dashed lines), and (iii) in
  setting $M_{FUV}^{*}$ at $z<0.2$ (dashed line) and reported as a
  reference in the $z>0.2$ panels (dotted lines).}
\label{FUV_LF_plot_separated} 
\end{figure}

\subsection{The FUV-band LFs using the different weighting
schemes}
\label{appendix_testing_weights}

To verify the robustness of our weights, we separately derive the
rest-frame FUV LF of the Deep and the Ultra-Deep surveys, using their
independent weighting schemes, as described in
Appendixes~\ref{appendix_weights_deep} and
~\ref{appendix_weights_udeep}.  Figure~\ref{FUV_LF_plot_separated}
shows the $1/V_{max}$ data points up to the respective LF bias limit
(see Sect.~\ref{lf_methods}) for the two surveys, together with the
STY LFs of the combined Deep and Ultra-Deep surveys (see
Sect.~\ref{lf_para} and Fig.~\ref{FUV_LF_plot}), using the merged
weighting scheme, as described in
Appendix~\ref{appendix_weights_merged}.  On one side, the LF data
points of the two surveys fully agree from $z=0.2$ to $z=3.5$ in
their common range of luminosities. This is an a-posteriori check of
the reliability of our individual weighting schemes. On the other
side, the STY LFs estimated with the merged sample perfectly overlap
the $1/V_{max}$ data points estimated with the two individual surveys.
This confirms the reliability of the weighting scheme applied to our
merged sample of individual surveys with different depth.


\section{Verifying the robustness of our rest-frame FUV-band LF}\label{LFallbands_app}

\begin{table*} 
  \caption{Schechter parameters ($M^{*}$, $\alpha$, $\phi^{*}$) and total Luminosity Density LD for
  the rest-frame FUV- and NUV-band galaxy luminosity function for the
  VVDS Deep+Ultra-Deep sample with the cosmology ($\Omega_{m}$,
  $\Omega_{\Lambda}$, $h$)~=~(0.3, 0.7, 0.7).}
\label{FUVNUV_LF_table} 
\centering 
\begin{tabular}{r c | c c c c| c c c c}
\hline\hline
 &  & \multicolumn{4}{c|}{FUV} & \multicolumn{4}{c}{NUV}  \\
 &  Redshift range~\tablefootmark{a} 	 & M$^{*}_{FUV}$ & $\alpha$ &  $\phi^{*}$  & lg(LD$_{uc}$)~\tablefootmark{b} & M$^{*}_{NUV}$ & $\alpha$ &  $\phi^{*}$  & lg(LD$_{uc}$)~\tablefootmark{b}\\    
 & & (AB~mag)  & &  ($10^{-3}$~Mpc$^{-3}$) & W/Hz/Mpc$^3$ & (AB~mag)  & &  ($10^{-3}$~Mpc$^{-3}$) & W/Hz/Mpc$^3$\\ 
\hline
 & & \multicolumn{8}{c}{Free Schechter parameters}  \\
\hline                        
           & $0.05< z \leq 0.2$     &     $-18.6^{+0.4}_{-0.6}$   &  $-1.10^{+0.06}_{-0.06}$	&    $5.36^{+1.57}_{-1.46}$ & 18.86$^{+0.25}_{-0.13}$ & $-18.5^{+0.3}_{-0.4}$   &  $-1.08^{+0.06}_{-0.06}$   &  $6.35^{+1.59}_{-1.47}$   &  18.90$^{+0.17}_{-0.11}$ \\
 $\bullet$ & $0.2< z \leq 0.4$      &     $-18.3^{+0.1}_{-0.2}$   &  $-1.17^{+0.05}_{-0.05}$	&    $6.91^{+1.02}_{-0.95}$ & 18.87$^{+0.03}_{-0.03}$ & $-18.3^{+0.1}_{-0.1}$   &  $-1.02^{+0.04}_{-0.04}$   &  $10.16^{+1.06}_{-1.01}$  &  18.97$^{+0.03}_{-0.03}$ \\
 $\bullet$ & $0.4< z \leq 0.6$      &     $-18.4^{+0.1}_{-0.1}$   &  $-1.07^{+0.07}_{-0.06}$	&    $6.60^{+0.91}_{-0.86}$ & 18.85$^{+0.02}_{-0.02}$ & $-18.7^{+0.1}_{-0.1}$   &  $-1.08^{+0.05}_{-0.05}$   &  $6.82^{+0.84}_{-0.79}$   &  18.99$^{+0.02}_{-0.02}$ \\
 $\bullet$ & $0.6< z \leq 0.8$      &     $-18.3^{+0.1}_{-0.1}$   &  $-0.90^{+0.08}_{-0.08}$	&    $9.53^{+0.99}_{-0.99}$ & 18.93$^{+0.02}_{-0.01}$ & $-18.8^{+0.1}_{-0.1}$   &  $-0.95^{+0.08}_{-0.08}$   &  $9.27^{+1.01}_{-1.01}$   &  19.12$^{+0.02}_{-0.02}$ \\
 $\bullet$ & $0.8< z \leq 1.0$      &     $-18.7^{+0.1}_{-0.1}$   &  $-0.85^{+0.10}_{-0.10}$	&    $9.01^{+0.94}_{-0.96}$ & 19.04$^{+0.02}_{-0.01}$ & $-19.2^{+0.1}_{-0.1}$   &  $-0.81^{+0.09}_{-0.08}$   &  $9.48^{+0.87}_{-0.88}$   &  19.27$^{+0.01}_{-0.01}$ \\
 $\bullet$ & $1.0< z \leq 1.2$      &     $-19.0^{+0.2}_{-0.2}$   &  $-0.91^{+0.16}_{-0.16}$	&    $7.43^{+1.08}_{-1.15}$ & 19.12$^{+0.04}_{-0.03}$ & $-19.6^{+0.1}_{-0.1}$   &  $-0.88^{+0.13}_{-0.12}$   &  $6.57^{+0.86}_{-0.89}$   &  19.30$^{+0.02}_{-0.02}$ \\
 $\bullet$ & $1.2< z \leq 1.7$      &     $-19.6^{+0.2}_{-0.2}$   &  $-1.09^{+0.23}_{-0.23}$	&    $4.10^{+0.77}_{-0.87}$ & 19.13$^{+0.12}_{-0.07}$ & $-20.2^{+0.1}_{-0.2}$   &  $-1.05^{+0.17}_{-0.16}$   &  $3.49^{+0.58}_{-0.61}$   &  19.28$^{+0.06}_{-0.04}$ \\
           & $1.7< z \leq 2.5$      &     $-20.7^{+0.4}_{-0.7}$   &  $-1.65^{+0.55}_{-0.53}$	&    $2.23^{+1.85}_{-1.56}$ & 19.70$^{+3.25}_{-0.36}$ & $-20.7^{+0.3}_{-0.4}$   &  $-1.16^{+0.47}_{-0.45}$   &  $3.73^{+1.29}_{-1.67}$   &  19.56$^{+0.52}_{-0.14}$ \\
           & $2.5< z \leq 3.5$      &     $-20.8^{+0.3}_{-0.3}$   &  $-0.63^{+0.57}_{-0.53}$	&    $1.72^{+0.19}_{-0.36}$ & 19.17$^{+0.25}_{-0.10}$ & $-21.4^{+0.3}_{-0.3}$   &  $-1.15^{+0.42}_{-0.40}$   &  $1.32^{+0.37}_{-0.47}$   &  19.37$^{+0.38}_{-0.13}$ \\
           & $3.5< z \leq 4.5$      &     $-$ & $-$ & $-$ & $-$ & $-$ & $-$ & $-$  & $-$   \\				       
\hline															       
 & & \multicolumn{8}{c}{$M^{*}$ constrained~\tablefootmark{c}, $\alpha$ and $\phi^{*}$ free}  \\			       
\hline															       
 $\bullet$ &  $0.05< z \leq 0.2$     &     $-18.12$		   &  $-1.05^{+0.04}_{-0.04}$  &    $7.00^{+0.44}_{-0.44}$  & 18.76$^{+0.04}_{-0.04}$  &  		    &	  &	&   \\
\hline															       
 & & \multicolumn{8}{c}{$\alpha$ constrained~\tablefootmark{d}, $M^{*}$ and $\phi^{*}$ free}  \\		 	       
\hline 	 										 				       
 $\bullet$ &  $1.7< z \leq 2.5$      &     $-20.4 ^{+0.1 }_{-0.1 }$   & $-1.30$ 	       &    $3.37^{+0.24}_{-0.24}$  &  19.46$^{+0.03}_{-0.03}$ &	&		 & &	 \\
 $\bullet$ &  $2.5< z \leq 3.5$      &     $-21.4 ^{+0.1 }_{-0.1 }$   & $-1.50$ 	       &    $0.86^{+0.05}_{-0.05}$  &  19.40$^{+0.02}_{-0.02}$ &	&		 & &	 \\
 $\bullet$ &  $3.5< z \leq 4.5$      &     $-22.2 ^{+0.2 }_{-0.2 }$   & $-1.73$ 	       &    $0.11^{+0.01}_{-0.01}$  &  19.10$^{+0.06}_{-0.05}$ &	&		 & &	 \\
           &  $1.7< z \leq 2.5$      &     $-20.3 ^{+0.1 }_{-0.1 }$   & $-1.10$ 	       &    $3.94^{+0.28}_{-0.28}$  &  19.41$^{+0.02}_{-0.02}$ &	&		 & &	 \\
           &  $2.5< z \leq 3.5$      &     $-21.1 ^{+0.1 }_{-0.1 }$   & $-1.10$ 	       &    $1.27^{+0.07}_{-0.07}$  &  19.24$^{+0.02}_{-0.01}$ &	&		 & &	 \\
           &  $3.5< z \leq 4.5$      &     $-21.8 ^{+0.1 }_{-0.1 }$   & $-1.10$ 	       &    $0.22^{+0.02}_{-0.02}$  &  18.73$^{+0.03}_{-0.03}$ &	&		 & &	 \\
           &  $1.7< z \leq 2.5$      &     $-20.8 ^{+0.2 }_{-0.2 }$   & $-1.73$ 	       &    $1.95^{+0.14}_{-0.14}$  &  19.79$^{+0.05}_{-0.04}$ &	&		 & &	 \\
           &  $2.5< z \leq 3.5$      &     $-21.6 ^{+0.1 }_{-0.1 }$   & $-1.73$ 	       &    $0.60^{+0.03}_{-0.03}$  &  19.62$^{+0.03}_{-0.03}$ &	&		 & &	 \\
\hline                                   
\end{tabular}

\tablefoot{
  \tablefoottext{a}The entries labelled with a filled dot are the
  chosen Schechter parameters taken for the rest-frame FUV LF used
  throughout the paper, and reported in Table~\ref{FUV_table_final}.
  \tablefoottext{b}  The quoted errors are those induced by the STY LF 
  fit; see Table~\ref{FUV_table_final} for errors including also 
  other sources of uncertainty. 
  \tablefoottext{c}$M^{*}$ is set at $z\leq0.2$ assuming the local
  rest-frame FUV value determined by \cite{wyder2005}.
  \tablefoottext{d}$\alpha$ is set with different values at
  $z>1.7$. First, following an evolution with redshift as described in
  Sect.~\ref{lf_para} ($\alpha=-1.3$, $-1.5$, $-1.73$ at
  $1.7<z\leq2.5$, $2.5<z\leq3.5$, $3.5<z\leq4.5$, respectively);
  secondly, set as two likely extreme non-evolving values
  ($\alpha=-1.1$ and $\alpha=-1.73$). We remind that $\alpha=-1.1$
  corresponds to what we find at $1.2<z\leq 1.7$ (where we can still
  determine $\alpha$ with a reasonable error), and $\alpha=-1.73$ is
  the value found by \cite{bouwens2007} at $z=3.8$. Contrarily to the
  rest-frame FUV LF, we were not interested in setting any Schechter
  parameters for the rest-frame NUV LF.}
\end{table*}

In Sect.~\ref{lf}, we have presented our VVDS rest-frame FUV-band LF
estimates from $z=0.05$ up to $z=4.5$ (see Fig.~\ref{FUV_LF_plot}),
and we have tabulated the corresponding {\it best} Schechter
parameters in Table~\ref{FUV_table_final}.  Figure~\ref{plot_LF_all}
summarises in one panel our best LF fits.  In this Appendix, we give
further details and tests we have performed to ensure the robustness
of our {\it best} rest-frame FUV LF, in particular the flatness of the
LF faint-end slope at $z\la 1$ and the low normalisation of the LF at
$z\ga 3$.  Table~\ref{FUVNUV_LF_table} gives our extensive
computations of the Schechter parameters with the STY method, leaving
the three Schechter parameters totally free. Obviously, we cannot
constrain $M^{*}$ at $z\leq 0.2$ nor $\alpha$ at $z>1.7$. We fixed
these two parameters as described in Sect.~\ref{lf_para}, and we also
report the results in Table~\ref{FUVNUV_LF_table}. The entries labelled
with a black dot are our final choices taken for our studies and also
reported in Table~\ref{FUV_table_final}.

As shown in Fig.~\ref{plot_LF_all}, our FUV faint-end slope is quite
flat ($\alpha\simeq-1$) at $z<1.7$, that is up to the highest redshift
where we can constrain it.  The compilation of the $\alpha$ values
found in the literature (see Table~\ref{FUV_alpha_table}) shows very
scattered values at $0\lesssim z \lesssim 6$. Nevertheless, it is a
notable point that $\alpha$ has been generally found steeper (when
estimated and not fixed) than our value. To strengthen the
validity of our results against possible biases, we report below our
LFs computed using photometric redshifts, using only starburst galaxies,
and computed in the rest-frame NUV-band.

\subsection{The rest-frame FUV LF with zphot}

\begin{figure} 
\centering
\includegraphics[width=\linewidth]{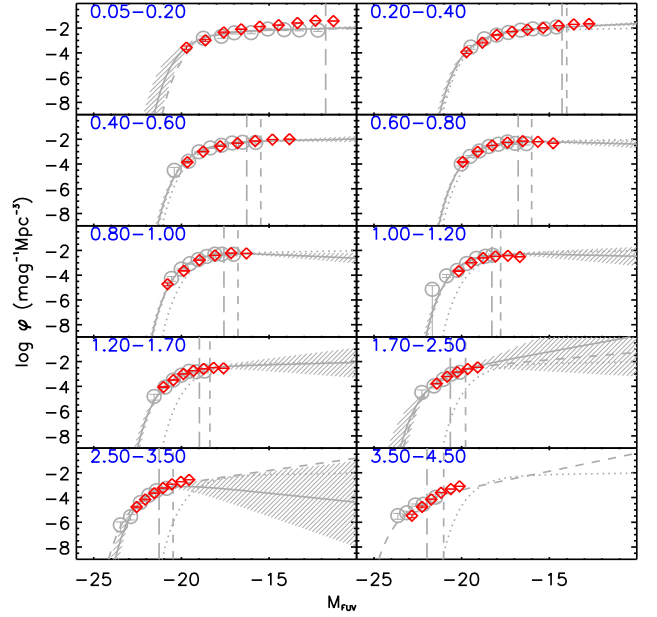}
\caption{Rest-frame FUV $1/V_{max}$ LF estimates from $z=0.05$ to
  $z=4.5$ derived with the VVDS $zphot$ catalogue up to $I_{AB}=26$
  (red open diamonds). They are plotted over a grey-scale version of
  Fig.~\ref{FUV_LF_plot}, that is the rest-frame FUV LF of the VVDS
  Deep+Ultra-Deep spectroscopic sample up to
  $I_{AB}=24.75$.} 
\label{FUV_LF_plot_zphot}
\end{figure}

We derive the VVDS rest-frame FUV $1/V_{max}$ estimates using very deep
photometric redshifts ($zphot$, see Appendix~\ref{appendix_weights}
for details). The $zphot$ catalogue is complete up to $I_{AB}=26$ and
it consists in $\sim43000$ photometric sources over the VVDS Deep area. In
this case, the weighting scheme is obviously not used.  The bias limit
has been computed in each $z$ bin as explained in
Sec.~\ref{lf_methods}, and at any $z$ it is clearly fainter than the
bias limit in the VVDS Deep+Ultra-Deep spectroscopic sample.

Results are shown in Fig.~\ref{FUV_LF_plot_zphot}, and globally
overlap our {\it best} rest-frame FUV-band LF determinations using the
VVDS Deep+Ultra-Deep spectroscopic sample up to $I_{AB} = 24.75$.
Thus, our spectroscopic sample (brighter $1.25$~mag in $I_{AB}$ than
the $zphot$ sample) does not significantly miss any faint galaxies,
which strengthens the reliability of our flat faint-end slope. One
small discrepancy occurs at $z\leq 0.2$, where the photometric
$1/V_{max}$ points have an higher normalisation with a slightly steeper
slope.  It is due to the well known degeneracy in the computation of
photometric redshifts, i.e., some objects end up with a wrong too low
($zphot\lesssim0.3$) photometric redshift (see
e.g. \citealp{ilbert2009}).  Furthermore, it confirms once more that
our weighting scheme is adequate. This strengthens also the
reliability of our low $\phi^{*}$ value at $3.5<z\leq4.5$, which is
neither caused by the cut at $I_{AB}=24.75$ nor by an incomplete
weighting scheme. We refer the reader to
Sec.~\ref{LF_z2_discussion} and to \cite{mccracken2003} for a
discussion about photometric completeness with respect to
$I_{AB}$-band magnitude and surface brightness.

\subsection{The rest-frame FUV LF of starburst galaxies}\label{app_LF_T4}

\begin{figure} 
\centering
\includegraphics[width=\linewidth]{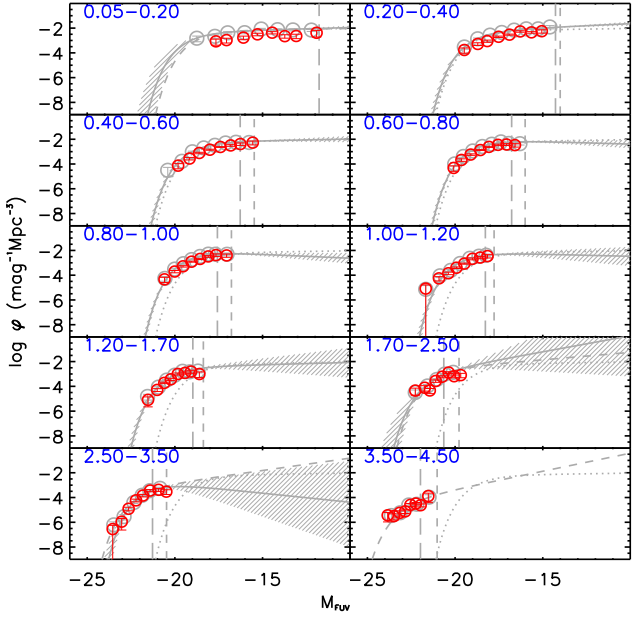}
\caption{Rest-frame FUV $1/V_{max}$ LF estimates from $z=0.05$ to
  $z=4.5$ for our intrinsically bluest galaxies (red open circles,
  VVDS Type~4, see text) in the VVDS Deep+Ultra-Deep spectroscopic
  sample. They are plotted over a grey-scale version of
  Fig.~\ref{FUV_LF_plot}, that is the rest-frame FUV LF of the total VVDS
  Deep+Ultra-Deep spectroscopic sample.}
\label{FUV_LF_plot_T4} 
\end{figure}

We are interested to check whether the flatness of our slope at
$z<1.7$ could be due to a very low density of intrinsically reddish
galaxies, which could have hidden, in the total sample, very high
densities of faint intrinsically blue galaxies. Indeed an $I$-band
selection does not select only starbursting galaxies, in contrast with
$UV$-band selected samples like GALEX \citep{arnouts2005}.

We have classified our galaxies of the VVDS Deep+Ultra-Deep
spectroscopic redshift sample according the VVDS scheme of four {\it
photometric types}, as described in \cite{zucca2006_VVDS_LF}.  Here,
we consider the `Type~4' galaxies, corresponding to galaxies for which
the best template fit over the $u^{*}g'r'i'z'JHK_{s}$ broad bands is
among the bluest galaxy templates, i.e., a starburst or irregular
galaxy template. The Type~4 rest-frame FUV-band $1/V_{max}$ LF estimates
are shown in Fig.~\ref{FUV_LF_plot_T4}. At $z<1.7$, their
normalisation is obviously lower with respect to the total sample,
because the fraction of Type~4 galaxies in our sample goes from
$\sim40$\% at $z\sim0.1$ up to $\sim85$\% at $z\sim4$ (when
considering galaxies brighter than the bias limit).  The faint-end LF
slope for Type~4 galaxies is very similar to the one found for the
total sample. Our flat slope is obviously not due to very low
densities of faint intrinsically red galaxies, which could have
overcompensate very high densities of faint intrinsically blue
galaxies. Thus, it reinforces the flatness of our slope.

\subsection{The rest-frame NUV LF}

\begin{figure}
\centering
\includegraphics[width=\linewidth]{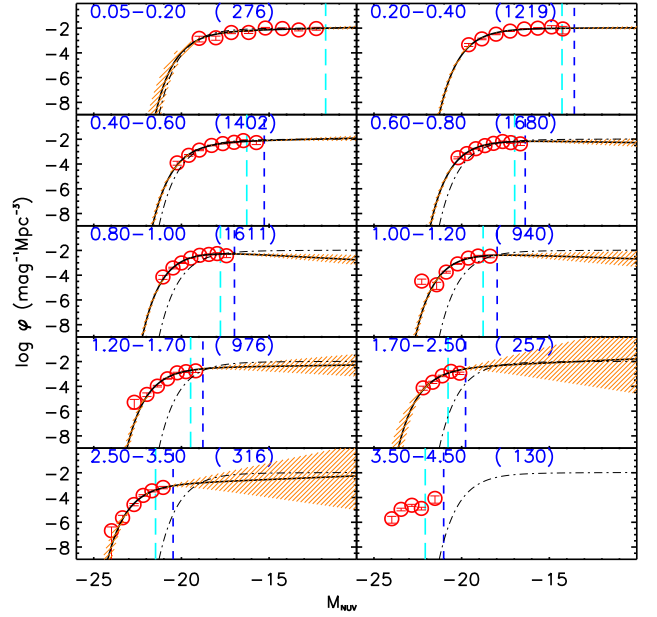}
\caption{Rest-frame NUV-band LFs from $z=0.05$ to $z=4.5$ of the VVDS
  Deep+Ultra-Deep sample, in 10 redshift bins as indicated in each
  panel.  Red circles represent the $1/V_{max}$ data
  points up to the LF bias limit and the corresponding number of
  galaxies are given in parenthesis. The vertical cyan long-dashed and
  blue short-dashed lines correspond to the LF bias limits for the
  Deep and Ultra-Deep surveys, respectively. The black solid curve and
  the associated shaded area is the STY LF estimate assuming free
  Schechter parameters and its associated error. The dot-dashed curve is
  the same in each panel, and it corresponds to the solid curve in
  the $0.2<z<0.4$ redshift bin, as reference.  All Schechter
  parameters are given in Table~\ref{FUVNUV_LF_table}.  For the
  NUV-band, we were not interested in fixing either $M^{*}$ nor
  $\alpha$. We note that in the last redshift bin the STY fit is
  unconstrained, so there is no solid curve overplotted.}
\label{LF_NUV_app} 
\end{figure}

The last point to test is the fact that our rest-frame FUV magnitudes
are based on template extrapolations at $z\le1$, which could have an
impact on the faint-end LF slope. Our bluest photometric information
is in the $u^{*}$-381 CFHTLS filter, and thus the rest-frame FUV
emission is directly observed from $z\ge1$. Nevertheless, the
rest-frame NUV emission is directly observed at $z\ge0.4$.  Thus we
report our results from the rest-frame NUV-band LF of the
Deep+Ultra-Deep spectroscopic sample. We list the Schechter parameters
derived with the STY method, in Table~\ref{FUVNUV_LF_table}, and plot
the LFs in Fig.~\ref{LF_NUV_app}. We note that we were not interested
in trying to fix $M{*}$ at low $z$, nor to fix $\alpha$ at very high
$z$. These estimates enable us to measure the faint-end slope at
$0.4\lesssim z \lesssim 1$, that is where the FUV intrinsic luminosity is
extrapolated but not the NUV intrinsic luminosity. Again, we do not
find a steeper faint-end slope in the rest-frame NUV-band LFs.  As our
aim is to derive the SFRD from the rest-frame ultraviolet continuum
spanning from FUV to the NUV using the relation of \cite{kennicutt1998},
we can affirm that the extrapolation of the FUV-band luminosities does  
not change our SFRD results since the faint-end slope stays flat.

\begin{table*} 
\caption{Non exhaustive list of the faint-end slope values, $\alpha$,
of the rest-frame FUV-band luminosity functions at $0<z<6$, as found
in the literature. Values are sorted according to the redshift. The
effective wavelength of the rest-frame band at which the LF was determined is
also given.}
\label{FUV_alpha_table} 
\centering 
\begin{tabular}{l l l l | l l l l}
\hline\hline 
Reference 	 &  $z$~\tablefootmark{a}  &  $\alpha$~\tablefootmark{b}  &  $\lambda$  &  Reference 	 &  $z$~\tablefootmark{a}  &  $\alpha$~\tablefootmark{b}    & $\lambda$ \\    
\hline
\cite{wyder2005}		          &   0.055     &   $-1.22 ^{+ 0.07 }_{-  0.07 }$   	& 1500 &  \cite{wilson2002_SFH}~\tablefootmark{d}	   &   1.35     &  $-1.5  $ & 2500 \\  
{\bf This work VVDS Deep~\tablefootmark{c}} &    0.125    &    $-1.05 ^{+ 0.04 }_{-  0.04 }$  	& 1500 &  \cite{gabasch2004_LF}			 &   1.36     & $-1.07 $ & 1500 \\
\cite{tresse2007} VVDS Deep~\tablefootmark{c} &    0.14     &   $-1.13^{+ 0.10}_{- 0.11} $  	& 1500 &  {\bf This work VVDS Deep+Ultra-Deep}        &   1.45     & $-1.09 ^{+ 0.23 }_{-  0.23 }$ & 1500 \\
\cite{sullivan2000_UVLF}~\tablefootmark{d} &   0.15     &   $-1.51 $  			 	& 2000 &  \cite{oesch2010_LF}  			 &   1.5      & $-1.86 ^{+ 0.48 }_{-  0.48 }$  & 1500 \\
\cite{treyer1998_SFR}~\tablefootmark{d}    &   0.15     &   $-1.62 $  			 	& 2000 &  \cite{tresse2007}				 &   1.55     & $-1.6  $  & 1500 \\
\cite{arnouts2005} GALEX-VVDS	           &    0.3      &   $-1.19 ^{+ 0.15 }_{-  0.15 }$   	& 1500 &  \cite{sawicky2006_LF}			 &   1.7      & $-0.81 ^{+ 0.21 }_{-  0.15 }$  & 1700 \\
{\bf This work VVDS Deep+Ultra-Deep}      &    0.3      &   $-1.17 ^{+ 0.05 }_{-  0.05 }$   	& 1500 &  \cite{connolly1997_SFH}~\tablefootmark{d}   &   1.75    & $-1.3  $  & 2800 \\
\cite{tresse2007}	                   &   0.3      &   $-1.6  $ 		         	& 1500 &  \cite{oesch2010_LF}  &    1.75     &     $-1.72 ^{+ 0.15 }_{-  0.15 }$  & 1500 \\
\cite{wilson2002_SFH}~\tablefootmark{d}	   &   0.35     &   $-1$  		 	 	& 2500 &  \cite{gabasch2004_LF}	    &	 1.88	  & $-1.07 $  & 1500 \\
\cite{wilson2002_SFH}~\tablefootmark{d}	   &   0.35     &   $-1.5  $  			 	& 2500 &  \cite{oesch2010_LF}  &    1.9      &     $-1.59 ^{+ 0.52 }_{-  0.52 }$ & 1500 \\
\cite{arnouts2005} GALEX-VVDS   	   &    0.5      &   $-1.55 ^{+ 0.21 }_{-  0.21 }$   	& 1500 &  \cite{arnouts2005}	     &    2		&   $-1.49 ^{+ 0.24 }_{-  0.24 }$		& 1500 \\
{\bf This work VVDS Deep+Ultra-Deep}       &    0.5      &   $-1.07 ^{+ 0.07 }_{-  0.06 }$   	& 1500 &  {\bf This work VVDS Deep+Ultra-Deep}       &    2.1      &	    $-1.3  $  & 1500 \\
\cite{tresse2007}	                   &   0.51      &   $-1.6  $  		  	 	& 1500 &  \cite{sawicky2006_LF}	    &	 2.2	  & $-1.20 ^{+ 0.24 }_{-  0.22 }$		& 1700 \\
\cite{gabasch2004_LF}	                   &   0.63      &   $-1.07 $  			 	& 1500 &  \cite{reddy_steidel2009}	&    2.3      &        $-1.73 ^{+ 0.07 }_{-  0.07 }$	& 1700 \\
\cite{tresse2007}	                   &   0.69      &   $-1.6  $  			 	& 1500 &  \cite{oesch2010_LF}  &    2.5      &        $-1.76 ^{+ 0.14 }_{-  0.14 }$			& 1500 \\
\cite{arnouts2005} GALEX-VVDS	           &    0.7      &   $-1.60 ^{+ 0.26 }_{-  0.26 }$   	& 1500 &  \cite{gabasch2004_LF}		    &	 2.53	  &   $-1.07 $  			& 1500 \\
\cite{cowie1999_LD}~\tablefootmark{d} 	   &   0.7     &   $-1   $  			 	& 2000 &  \cite{arnouts2005}	    &	 2.9	  &   $-1.47 ^{+ 0.21 }_{-  0.21 }$		& 1500 \\
\cite{cowie1999_LD}~\tablefootmark{d} 	   &   0.7     &   $-1.5  $  			 	& 2000 &  {\bf This work VVDS Deep+ultra-Deep} 	&    3       &   $-1.5  $					& 1500 \\
{\bf This work VVDS Deep+Ultra-Deep}       &    0.7      &   $-0.90 ^{+ 0.08 }_{-  0.08 }$   	& 1500 &  \cite{sawicky2006_LF}		    &	 3	  &   $-1.43 ^{+ 0.17 }_{-  0.09 }$	& 1700 \\
\cite{connolly1997_SFH}~\tablefootmark{d}  &   0.75     &   $-1.3  $  			 	& 2800 &  \cite{steidel99}	&   3.04     &   $-1.6  $					& 1700 \\
\cite{oesch2010_LF}	                   &    0.75     &   $-1.54 ^{+ 0.26 }_{-  0.26 }$   	& 1500 &  \cite{tresse2007} VVDS Deep  &   3.04    &	$-1.6  $					& 1500 \\
\cite{wilson2002_SFH}~\tablefootmark{d}	   &   0.8     &   $-1   $  			 	& 2500 &  \cite{reddy_steidel2009}	&    3.05     &   $-1.73 ^{+ 0.13 }_{-  0.13 }$ 	& 1700 \\
\cite{wilson2002_SFH}~\tablefootmark{d}	   &   0.8     &   $-1.5  $  			 	& 2500 &  \cite{gabasch2004_LF}		  &    3.46	&   $-1.07 $				& 1500 \\
{\bf This work VVDS Deep+Ultra-Deep}       &   0.9       &   $-0.85 ^{+ 0.10 }_{-  0.10 }$   	& 1500 &  \cite{paltani2007}  VVDS Deep  &   3.5	&   $-1.4  $					& 1700 \\
\cite{tresse2007}  VVDS Deep	           &   0.9	 &  $-1.6  $			 	& 1500 &  \cite{tresse2007} VVDS Deep &   3.6      &	$-1.6  $					& 1500 \\
\cite{gabasch2004_LF}	                   &   0.96	 & $-1.07 $			 	& 1500 &  \cite{bouwens2007}	&    3.8      &   $-1.73 ^{+ 0.05 }_{-  0.05 }$ 		& 1600 \\
\cite{arnouts2005} GALEX-VVDS              &   1	 &  $-1.63 ^{+ 0.45 }_{-  0.45 }$	& 1500 &  {\bf This work VVDS Deep+Ultra-Deep} 	&    4       &   $-1.73 $					& 1500 \\
\cite{tresse2007} VVDS Deep	           &   1.09      &  $-1.6  $			 	& 1500 &  \cite{sawicky2006_LF}		&    4       &   $-1.26 ^{+ 0.40 }_{-  0.36 }$  	& 1700 \\
{\bf This work VVDS Deep+Ultra-Deep}       &    1.1      &  $-0.91 ^{+ 0.16 }_{-  0.16 }$	& 1500 &  \cite{steidel99}	&   4.13     &   $-1.6  $					& 1700 \\
\cite{connolly1997_SFH}~\tablefootmark{d}  &   1.25     &  $-1.3  $		         	& 2800 &  \cite{tresse2007} VVDS Deep  &   4.26      &   $-1.6  $					& 1500 \\
\cite{cowie1999_LD}~\tablefootmark{d}	   &   1.25	 &  $-1   $	    	         	& 2000 &  \cite{gabasch2004_LF}		&    4.51     &   $-1.07 $				& 1500 \\
\cite{cowie1999_LD}~\tablefootmark{d}	   &   1.25	 &  $-1.5  $			 	& 2000 &  \cite{bouwens2007}	&    5        &   $-1.66 ^{+ 0.09 }_{-  0.09 }$ 		& 1600 \\
\cite{oesch2010_LF}	                   &    1.25     &  $-1.76 ^{+ 0.23 }_{-  0.23 }$	& 1500 &  \cite{iwata2007}	&   5	   &   $-1.48 ^{+ 0.38 }_{-  0.32 }$		& 1700 \\
\cite{tresse2007} VVDS Deep	           &   1.29      &  $-1.6  $		         	& 1500 &  \cite{bouwens2007}	&    5.9      &   $-1.74 ^{+ 0.16 }_{-  0.16 }$ 		& 1350 \\
\cite{wilson2002_SFH}~\tablefootmark{d}	   &   1.35     &  $-1   $		         	& 2500 &  	&	       &   &  \\
\hline 
\end{tabular} 

\tablefoot{\tablefoottext{a}Mean or median redshift as quoted in each
work, or, if not specified, centre of the studied redshift interval.
  \tablefoottext{b}Faint-end slope of the UV LF, with error bars when
  it was estimated, with no error bars when it was fixed.
  \tablefoottext{c}Faint-end slope of the nearby FUV LF estimated after
  setting $M^{*}_{FUV}$.
  \tablefoottext{d}Value retrieved from the compilation of
  \cite{hopkins2004}.}
\end{table*}


\section{FUV number counts and luminosity functions at
$z\sim3$}\label{lf_counts_z3}

\begin{figure*} 
\centering
\includegraphics[width=0.4\linewidth]{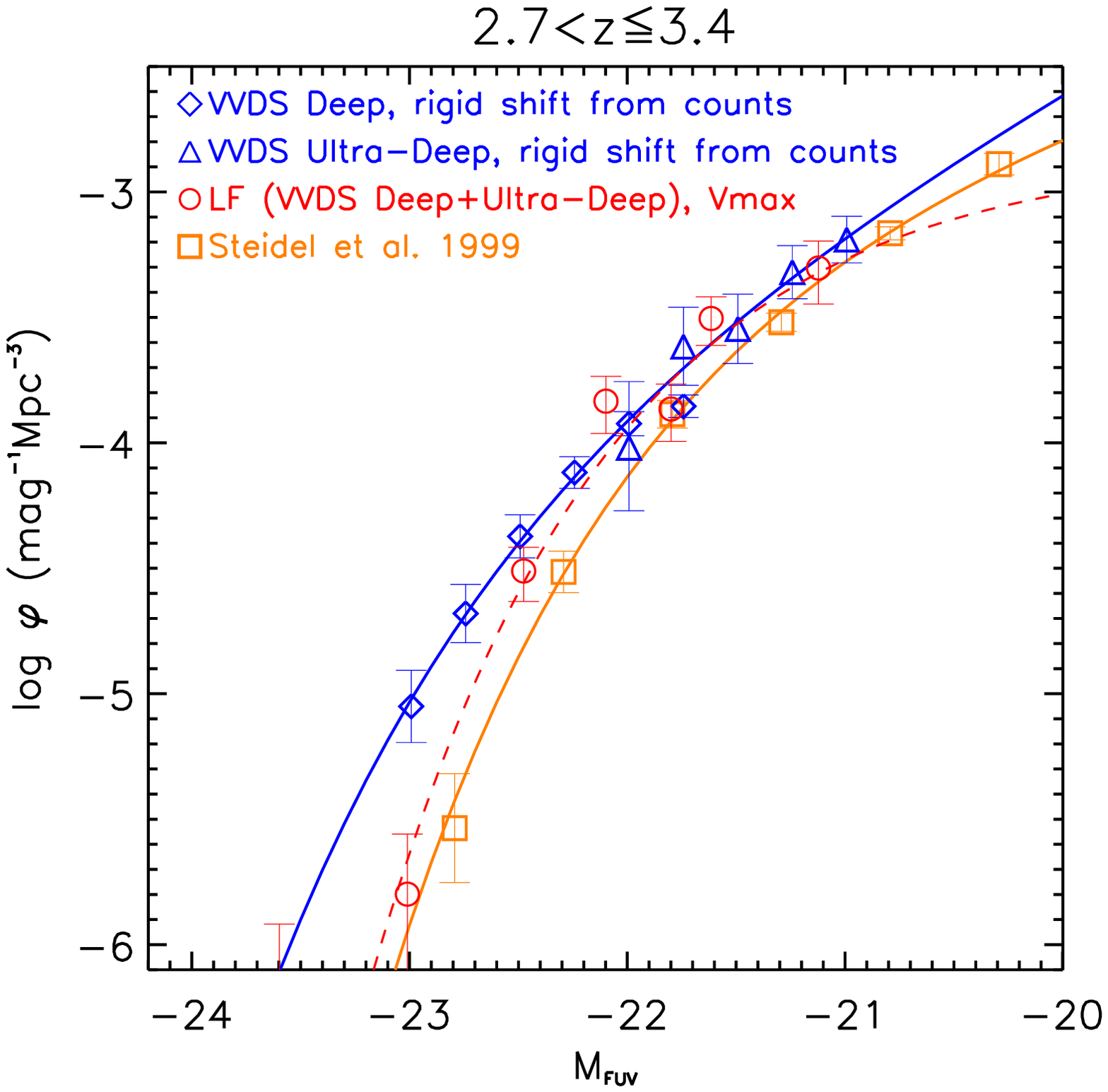}
\includegraphics[width=0.4\linewidth]{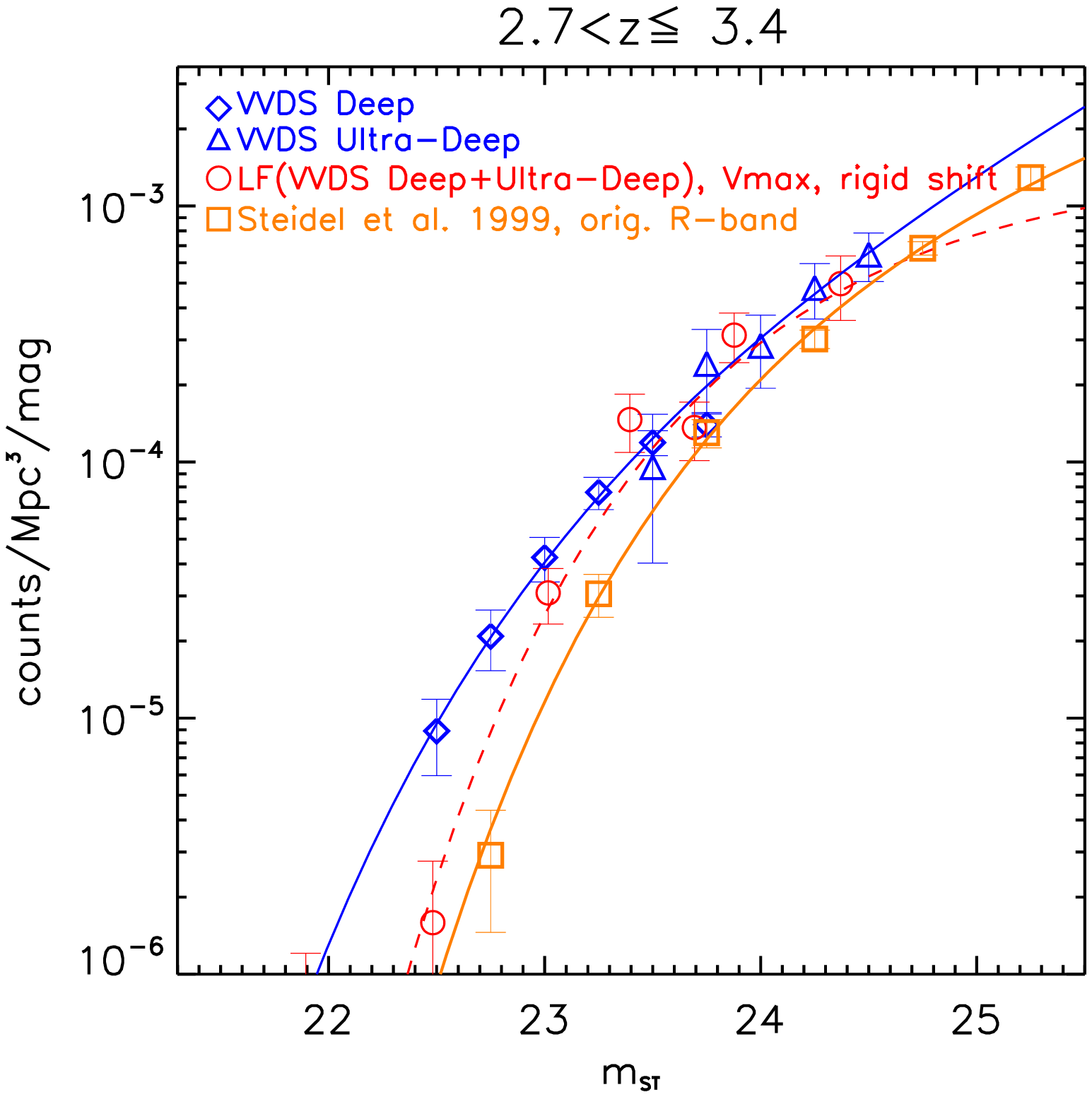}
\caption{{\it Left panel}. Comoving luminosity densities as a function
  of FUV (1700 $\AA$) absolute luminosities at $2.7 \leq z \leq 3.4$: $1/V_{max}$
  estimates of our VVDS Deep+Ultra-Deep sample fitted with a Schechter
  function (red circles and dashed curve, respectively); our
  $m_{ST}=(r+i)/2$ number counts per volume density for the Deep and
  Ultra-Deep surveys converted into FUV absolute magnitudes with a
  rigid shift (see text), and fitted with a Schechter function (blue
  diamonds and triangles, and solid line, respectively); the
  $m_{ST}=\mathcal{R}$ number counts per volume density of
  \cite{steidel99} converted into FUV absolute magnitudes with their
  rigid shift (see text), and fitted with a Schechter function by them
  (orange squares and solid line, respectively).  {\it Right
    panel}. Number counts per unit volume of galaxies as a function of
  $m_{ST}$ apparent magnitudes at $2.7 \leq z \leq 3.4$: our number
  counts from the VVDS Deep and Ultra-Deep surveys (blue diamonds and
  triangles, respectively); our VVDS number counts derived from our
  VVDS $1/V_{max}$ LF (see right panel) shifted rigidly in magnitude
  (red circles, see text); number counts from \cite{steidel99} (orange
  squares; see text
  for conversion details). The three curves correspond to the same
  curves in the left panel, shifted rigidly (see text). Note that the
  CFHTLS $(r+i)/2$ photometric system mimics the $\mathcal{R}$ one
  used by \cite{steidel99}.}
\label{FUV_LF_steidel_appendix} 
\end{figure*}

At $2.7\leq z \leq 3.4$, \cite{lefevre2005nat} and  Le~F\`evre et al. (in prep.)
show that the VVDS presents number counts per 0.5~mag interval of
$I_{AB}$ apparent magnitude per unit surface area (i.e.,
dN/0.5mag/arcmin$^2$) at least 2 times larger than those quoted by
\cite{steidel99} within the same redshift range. This is particularly
evident for bright galaxies (I$_{AB}\lesssim 23.5$). In contrast, the
VVDS rest-frame FUV LF is only $\sim50$\% higher than the one found by
\cite{steidel99}, as illustrated in the left panel of our 
Fig.~\ref{FUV_LF_steidel_appendix}, and this holds
only for $M_{FUV}<-22.5$. Here, we demonstrate that these
two results are not in conflict. We remark that throughout the paper we compute our
VVDS FUV absolute magnitudes using a filter centred at 1500 $\AA$,
while \cite{steidel99} use a FUV-band filter centred at 1700 $\AA$. To
compare our results with their work, in this section we compute our
VVDS FUV absolute magnitudes using the same filter as
\cite{steidel99}, even though our VVDS FUV-150 and VVDS-1700 LFs are
very similar.

As \cite{steidel99} begin their analysis from galaxy number counts as
a function of $\mathcal{R}$ apparent magnitude, we overplot the same
for the VVDS Deep and Ultra-Deep surveys in the right panel of
Fig.~\ref{FUV_LF_steidel_appendix}. We remark that in this Figure we
plot the number of galaxies per unit magnitude and per unit volume
(i.e., dN/mag/Mpc$^3$), while \cite{steidel99} cite in their Table~3
the surface density number counts for 0.5 mag intervals
(dN/0.5mag/arcmin$^2$). To convert their values in dN/mag/Mpc$^3$
units, we multiplied them $\times 2$ to obtain the correct magnitude
interval, and we divided for the effective volume of each arcmin$^2$,
quoted in their Table~3 (with the same cosmology that we adopt,
$\Omega_m=0.3$, $\Omega_{\Lambda}=0.7$). Our number counts are
weighted as described in Appendix~\ref{appendix_weights_deep} and
\ref{appendix_weights_udeep}, and they are realized as a function of
$m_{ST}=(r+i)/2$ apparent magnitude, i.e., the mean value of the $r$-
and $i$-CFHTLS broadband filters, because it mimics the
$\mathcal{R}$-broadband filter used by \cite{steidel99}.  We observe
that the VVDS counts are clearly higher than those from
\cite{steidel99}, at least at the brightest magnitudes.

Next, from their number counts, \cite{steidel99} derive their
rest-frame FUV LF (shown in the left panel of
Fig.~\ref{FUV_LF_steidel_appendix}), translating apparent
magnitudes into absolute magnitudes with the following rigid shift:

\begin{equation} \displaystyle 
M_{FUV} = m_{\mathcal{R}}-DM(z_{med})+2.5\log(z_{med}+1), 
\label{shift_steidel} 
\end{equation}

where DM($z_{med}$) is the distance modulus defined as
$5\times\log(D_{L}(z_{med})/10{\rm pc})$, with $D_{L}(z_{med})$ the
luminosity distance at $z_{med}$, and $z_{med}$ the median redshift of
the studied sample. For their sample at $2.7\leq z \leq 3.4$
($z_{med}=3.04$), it gives:

\begin{equation} \displaystyle
M_{FUV} = m_{\mathcal{R}} - 45.55.
\label{shift_steidel_value} 
\end{equation}

For the VVDS Deep+Ultra-Deep sample, at $2.7\leq z \leq 3.4$ we have
$z_{med}=2.92$. With the same method, starting from our observed
number counts (see right panel of Fig.~\ref{FUV_LF_steidel_appendix}),
we obtain the rest-frame FUV LF data points, as illustrated in the
left panel of Fig.~\ref{FUV_LF_steidel_appendix}, and we fitted them with a
Schechter function. The result is that our $dN$-derived LF is higher
at the bright end than our direct $1/V_{max}$ LF data points, as derived
in Sect.~3.2, and here fitted also with a Schechter function.  We
conclude that our $dN$-derived LF presents a different shape than our
$1/V_{max}$-derived LF.

Below we verify that, as expected, it is not correct to use a simple, rigid shift to
transform our apparent magnitudes to absolute ones, as done with
Eq.~\ref{shift_steidel}, to obtain correct $dN$-derived LFs. The main
reasons are the following.

\begin{figure} 
\centering
\includegraphics[width=\linewidth]{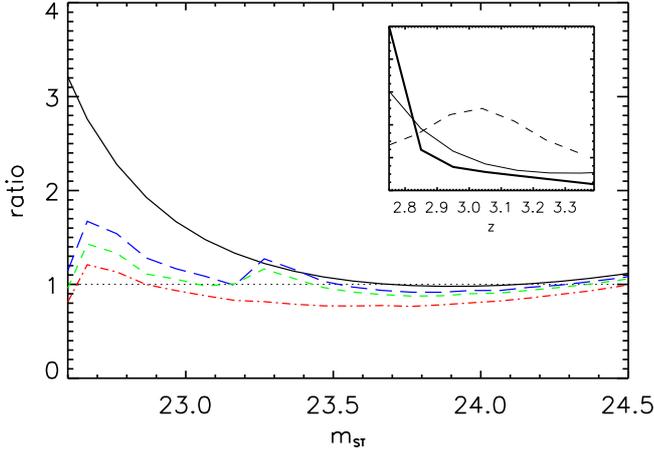}
\caption{As a function of $m_{ST}$, ratio of the directly observed VVDS FUV
number counts at $2.7\leq z \leq 3.4$ to the number counts derived in
various ways from the FUV luminosity function in the same $z$
range. The black dotted line (ratio=1) is the ideal case where both
counts are identical. The black solid line represents the ratio of the
observed VVDS to the VVDS LF derived counts using
Eq.~\ref{shift_steidel}, $n(m_{ST}, LF)$ (i.e. blue solid line to the
red dashed line of Fig.~\ref{FUV_LF_steidel_appendix}, right panel).
The other lines represent the ratio of the observed VVDS number counts
(the above mentioned blue line) to the following various simulated
distributions: $n(m_{ST},z)$ (blue long-dash curve),
$n(m_{ST},z,K,col)$ (green short-dash curve) and $n(m_{ST}, \Delta M)$
(red dot-dash curve). See text for details about these simulated
distributions.  In the inset, we plot the global shape of the redshift
distributions in \cite{steidel99} (dashed line) and in our work,
separately for faint and bright galaxies (thin and thick solid lines,
respectively). The normalisation on the $y$-axis is consistent between
the two solid lines, while the normalisation of the dashed line has
been arbitrarily chosen to better illustrate the different shapes.}
\label{ratio_counts_LF} 
\end{figure}

\begin{description}
\item[1) {\it The $n(z)$ shape.}] In the range $2.7\leq z \leq 3.4$, 
  the redshift distribution $n(z)$ of an $I$-band selected survey as
  observed by the VVDS decreases as a function of $z$ as shown in
  Le~F\`evre et al. (in prep.), in particular in a faster way for
  brighter galaxies. In contrast, due to the colour selection function
  used to find Lyman-break galaxies, the $n(z)$ in \cite{steidel99}
  peaks around its median value within the same redshift range. The
  shapes of the different $n(z)$ are shown in the inset of
  Fig.~\ref{ratio_counts_LF}. If the $n(z)$ is skewed, it is crucial
  to use the correct redshift distribution in Eq.~\ref{shift_steidel},
  and not simply the median redshift value.
\item[2) {\it The $K$- and colour corrections.}] \cite{steidel99}
  restrict the ($K$+colour) term to $2.5log(1+z)$, as in
  Eq.~\ref{shift_steidel}, because they work with the observed
  $\mathcal{R}$-band, corresponding to the FUV light emitted at
  $z\sim3$.  Nevertheless, this is correct for $z\sim3.0$ galaxies
  only.

\end{description}

We made a simple simulation to illustrate these two aspects. We
started with a set of $3\times10^6$ galaxies distributed in FUV
absolute magnitude like the Schechter fit of our $1/V_{max}$-derived FUV
LF. We want to know whether, with a selection function as the VVDS, we
recover our observed galaxy number counts as a function of $m_{ST}$
apparent magnitude at $2.7\leq z \leq 3.4$. Using
Eq.~\ref{shift_steidel}, we operate a rigid shift to transform our FUV
absolute magnitudes into $m_{ST}$ apparent magnitudes, and to obtain a
LF-derived number counts as illustrated in the right panel of
Fig.~\ref{FUV_LF_steidel_appendix}. We call this LF-derived
distribution $n(m_{ST},LF)$. We see that these LF-derived number
counts are not in agreement with our observed number counts which are
at least twice higher at $m_{ST}<23$.  We illustrate in
Fig.~\ref{ratio_counts_LF} the ratio of our observed number counts to
our $n(m_{ST},LF)$ as a function of $m_{ST}$.

We will now test the $n(z)$ and the ($K$+colour) corrections,
respectively.  We fit separately the $n(z)$ of VVDS galaxies with
$m_{ST}\leq 23$ and with $m_{ST}>23$, to account for the steeper
$n(z)$ decrease for bright magnitudes (see inset in
Fig.~\ref{ratio_counts_LF}).  Given the monotonic (even if scattered)
relation between observed and absolute magnitudes, $m_{ST}\sim 23$
corresponds to $M_{FUV}\sim -21.1$. Thus, for each previously
simulated galaxy, we assign a redshift in respecting the different
$n(z)$ distributions for galaxies brighter or fainter than $M_{FUV} =
-21.1$, that is we do not anymore assume a single median redshift.  We now
compute the observed $m_{ST}$ magnitudes of our simulated sample using
Eq.~\ref{shift_steidel}, where we substitute $z_{med}$ with the
particular redshift that has been assigned to each galaxy. We obtain
a $m_{ST}$ distribution that we call $n(m_{ST},z)$.  The ratio of our
observed number counts to the $n(m_{ST},z)$ is shown in
Fig.~\ref{ratio_counts_LF}. The $n(m_{ST},z)$ distribution is now
closer to our observed number counts than the $n(m_{ST},LF)$
distribution.

As a further step, we account for the ($K$+colour) terms for each
previously simulated galaxy, that is:

\begin{equation} \displaystyle 
M_{FUV} = m_{ST}-DM(z)-K_{ST}(z) - (m_{ST} - m_{FUV})^{(SED,z=0)},
\label{mag_abs_eq} 
\end{equation}

where $K_{ST}(z)$ is the $K$-correction at $z$ and $(m_{ST} -
m_{FUV})$ is the colour term at $z=0$, both derived from the best SED
fitting template.  We have modelled the best fitting template
distribution in the VVDS sample, which gives us the VVDS distributions
of $K_{ST}(z)$ and $(m_{ST} - m_{FUV})^{(SED,z=0)}$ terms. Also in
this case we assumed the two $n(z)$ distributions (for faint and
bright galaxies) described above.  Using the specific distribution of
all the terms in Eq.~\ref{mag_abs_eq}, we computed a new set of
$m_{ST}$ starting from our sample of absolute magnitudes, and obtained
their distribution ($n(m_{ST},z,K,col)$).  We show in
Fig.~\ref{ratio_counts_LF} the ratio of our observed number counts to
the $n(m_{ST},z,K,col)$ distribution.
We observe that the modelisation of the $K$+colour term further
improves the match with the observed number counts. Still, the ratio
is not equal to unity, which demonstrates 
that the reality is more complex than a simple simulated recipe.

To mimic the VVDS observed data, but still with a very simple test, we
have fitted with a Gaussian function the distribution of $\Delta M =
M_{FUV}-m_{ST}$ in the VVDS sample.  Subtracting from each simulated
$M_{FUV}$ a value of $\Delta M$ extracted randomly from the
distribution, we obtained a new $m_{ST}$ distribution, that we called
$n(m_{ST}, \Delta M)$. It is clear that this last distribution (see
Fig.~\ref{ratio_counts_LF}) mimics better than the others the real
observed number counts distribution at bright magnitudes, but then the
match for fainter galaxies is slightly worse.

In summary, this simple exercise confirms that it is very dangerous to
transform number counts within a given redshift range to a luminosity
function in the case of a skewed $n(z)$ within the redshift interval
considered, and/or in the case of dissimilar $n(z)$ for different
galaxy populations. The fact that the $n(z)$ of colour-selected
samples is very different from the $n(z)$ of magnitude limited
samples, implies that number counts when using colour-selected samples
are not representative of the number counts for the complete galaxy
population. We refer the reader to \cite{lefevre2005nat} and
Le~F\`evre et al. (in prep.) for more details on number counts and
sample selections.

\end{document}